The Alan Turing Institute

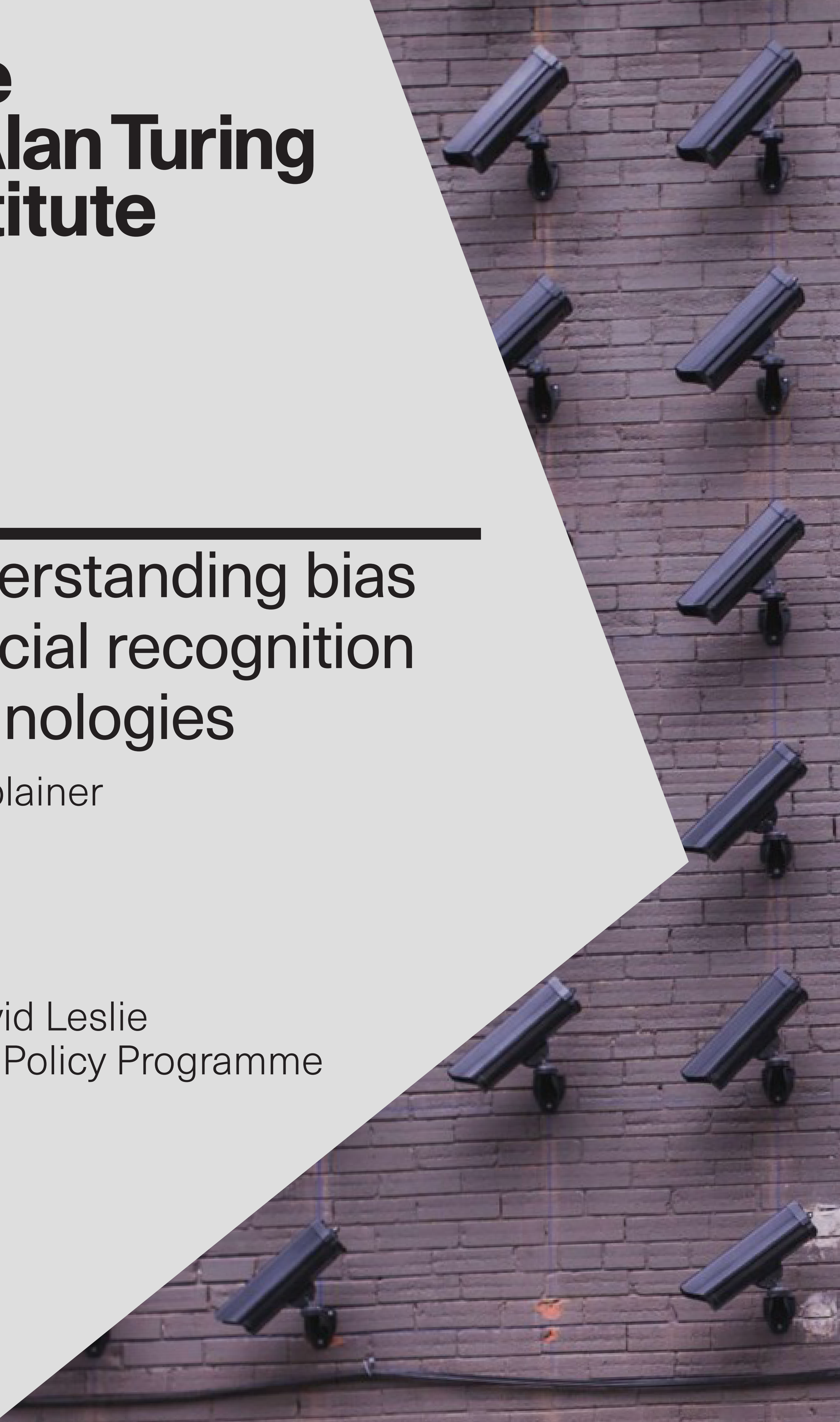

# Understanding bias in facial recognition technologies

## An explainer

Dr David Leslie
Public Policy Programme

The Alan Turing Institute

The Public Policy Programme at The Alan Turing Institute was set up in May 2018 with the aim of developing research, tools, and techniques that help governments innovate with data-intensive technologies and improve the quality of people's lives. We work alongside policy makers to explore how data science and artificial intelligence can inform public policy and improve the provision of public services. We believe that governments can reap the benefits of these technologies only if they make considerations of ethics and safety a first priority.

Please note, that this explainer is a living document that will evolve and improve with input from users, affected stakeholders, and interested parties. We need your participation. Please share feedback with us at policy@turing.ac.uk This research was supported, in part, by a grant from ESRC (ES/T007354/1) and from the public funds that make the Turing's Public Policy Programme possible. https://www.turing.ac.uk/research/research-programmes/public-policy

---

This work is dedicated to the memory of Sarah Leyrer. An unparalled champion of workers' rights and equal justice for all. A fierce advocate for the equitable treatment of immigrants and marginalised communities. An unwaivering defender of human dignity and decency.

---

# Table of Contents

# Introduction

Over the past couple of years, the growing debate around automated facial recognition has reached a boiling point. As developers have continued to swiftly expand the scope of these kinds of technologies into an almost unbounded range of applications, an increasingly strident chorus of critical voices has sounded concerns about the injurious effects of the proliferation of such systems on impacted individuals and communities. Not only, critics argue, does the irresponsible design and use of facial detection and recognition technologies (FDRTs)[1] threaten to violate civil liberties, infringe on basic human rights and further entrench structural racism and systemic marginalisation, but the gradual creep of face surveillance infrastructures into every domain of lived experience may eventually eradicate the modern democratic forms of life that have long provided cherished means to individual flourishing, social solidarity and human self-creation.

Defenders, by contrast, emphasise the gains in public safety, security and efficiency that digitally streamlined capacities for facial identification, identity verification and trait characterisation may bring. These proponents point to potential real-world benefits like the added security of facial recognition enhanced border control, the increased efficacy of missing children or criminal suspect searches that are driven by the application of brute force facial analysis to largescale databases and the many added conveniences of facial verification in the business of everyday life.

Whatever side of the debate on which one lands, it would appear that FDRTs are here to stay. Whether one is unlocking an iPhone with a glance, passing through an automated passport checkpoint at an airport, being algorithmically tagged in photos on social media, being face-scanned for professional suitability by emotion detection software at a job interview or being checked against a watchlist database for suspiciousness when entering a concert hall or sporting venue, the widening presence of facial detection and analysis systems is undeniable. Such technologies are, for better or worse, ever more shaping the practices and norms of our daily lives and becoming an increasingly integrated part of the connected social world. Indeed, it is, as a consequence of this, easy to succumb to the sneaking suspicion that the coming pervasiveness of facial analysis infrastructures is all but unavoidable.

The seeming inevitability of this sociotechnical state of play is, however, problematic for at least three reasons. First, it has created a fraught point of departure for diligent, proactive and forward-looking considerations of the ethical challenges surrounding FDRTs. Rather than getting out ahead of their development and reflecting in an anticipatory way about their potentially harmful or discriminatory impacts, much of the discussion around the ethics of automated facial recognition has taken the existence of such technologies for granted as a necessary means for securing public safety or improving efficiency. Such conversations have focused instead on how to right the things that are going wrong with present practices of designing and deploying them. They have concentrated on how to remedy the real-world harms being done.

As critical as this task may be, the focus on remediation has meant that more basic ethical concerns surrounding the very justifiability and ethical permissibility of the use of FDRTs have tended to remain in the shadows—with concerns that broach the transformative effects of the spread of these technologies on individual self-development, democratic agency, social cohesion, interpersonal intimacy and community wellbeing often being depreciated, set aside or shelved altogether. This has led to a troubling absence in public discourse of the widespread engagement of fundamental moral questions such as: Should we be doing this in the first place? Are these technologies ethically permissible to pursue given the potential short- and long-term consequences of their broad-scale development? Do technologists and innovators in this space stand on solid ethical ground in flooding society, whatever the cost, with these expanding capacities to globally automate unbounded personal identification and ubiquitous smart surveillance?

The second problem with the outwardly inevitable proliferation of FDRTs is that, unlike the deterministic laws of the Newtonian universe, the inevitability itself has not operated evenly and uniformly. The march forward of facial analysis technologies has not been neutral with regard to the distribution of its harms and benefits. Rather, it has trampled over the rights and freedoms of some all while generating windfalls of profit, prestige and convenience for others. Throughout the evolution of FDRTs, from the very first innovations in data-driven facial detection in the early 2000s to the churning architectures of the convolutional neural networks that power facial recognition today, bias and discrimination have been as much a part of the development and use of these technologies as pixels, parameters and data have. Telling this difficult and disconcerting story will be the primary purpose of this explainer.

Briefly, the tale of bias and discrimination in FDRTs actually begins in the 19th century with the development of photography. For generations, the chemical make-up of film was designed to be best at capturing light skin. Colour film was insensitive to the wide range of non-white skin types and often failed to show the detail of darker-skinned faces. Deep-seated biases toward this privileging of light skin as the global norm of ideal flesh tone endured into the era of digital cameras and eventually found their way into the software behind automated facial recognition systems. As these technologies became primarily data-driven, their dependence on largescale datasets spurred new forms of bias. Imbalanced datasets with less representation of marginalised demographic groups would train FDRTs to be less accurate for them, and, up to very recently, pretty much all the available largescale face datasets were over-representative of white males and under-representative of people of colour and women. Such an entrenchment of systemic discrimination has also cropped up in the labelling and annotating of datasets where categories of “race,” “ethnicity,” and “gender” are unstable and reflect cultural norms and subjective categorisations that can lead to forms of scientific racism and prejudice. Beyond this, technical and design issues that make FRDTs perform differently for dominant and subordinated groups are only the beginning of the story that must be told of how structural injustices and systemic discrimination may be exacerbated by the widescale use of these kinds of biometric technologies.

The third problem with the seeming inevitability of the spread of FDRTs builds on the first two. It has to do with the fact that, as recent history has demonstrated, the forward march of these technologies is not inevitable at all. Current events show that members of contemporary digital society are starting to militate against the intolerable forms of racialisation, discrimination and representational and distributional injustice which are surfacing in the development and deployment of FDRTs. Critically-minded people from both tech and non-tech commnunities are not only  doing this by calling into question the justifiability of these systems, but they are also showing themselves to be capable of pumping the brakes as a consequence these qualms as well. As recent waves of corporate back-peddling, Big Tech moratoria, successful litigations and local facial recognition bans have demonstrated, the critical force of a sort of retrospective anticipation that responds to societal harms by revisiting the legitimacy and justifiability of FDRTs is playing a major role in reining them in.

Taken together, these various high-profile course corrections intimate that society is still capable of reclaiming its rightful place at the controls of technological change. Though not optimal, the capacity to challenge the ethical justifiability of FDRTs through retrospective anticipation might as yet be a powerful tool for spurring more responsible approaches to the rectification, reform and governance of these technologies. When community-led, deliberatively-based and democratically-shepherded, such critical processes can also be a constructive force that recovers a degree and manner of participatory agency in socio-technical practices that have gone awry.

### *The plan of the explainer*

In what follows, I will take a deeper dive into all of these issues. I will begin by providing a brief primer on the technical dimension of facial detection and analysis technologies. I will then explain some of the factors that lead to biases in their design and deployment, focusing on the history of discrimination in the development of visual representation technologies. Next, I will take a step back and broach the wider ethical issues surrounding the design and use of FDRTs—concentrating, in particular, on the distributional and recognitional injustices that arise from the implementation of biased FDRTs. I will also describe how certain complacent attitudes of innovators and users toward redressing these harms raise serious concerns about expanding future adoption.  The explainer will conclude with an exploration of broader ethical questions around the potential proliferation of pervasive face-based surveillance infrastructures and make some recommendations for cultivating more responsible approaches to the development and governance of these technologies.

## Background

To understand how bias can crop up in face detection and analysis algorithms, it would be helpful first to understand a little bit about the history of how computers came to be able to detect, analyse and recognise faces. This story really begins in the 1990s, when digital cameras started to commercially replace older film-based methods of photography. Whereas traditional cameras produced photos using celluloid film coated with light-sensitive chemicals, digital cameras used photosensors to capture images by converting light waves into grids of pixel values that recorded the patterns of brightness and colour generated by these incoming waves. The ability to record images electronically as an array of numerical values was revolutionary. It meant that photographs became automatically computer-readable (as patterns of numbers) and that they could be immediately stored and retrieved as data.

Around this same time, the internet began rapidly to expand. And, as more and more people started to use the web as a way to share and retrieve information, more and more digital images came to populate webpages, news sites and social media platforms like MySpace, Facebook and Flickr. Not only did this exponential increase in content-sharing usher in the big data revolution, the growing online reservoir of digital images provided a basis for the development of the data-driven computer vision technologies that have evolved into today's FDRTs (Figure 1). The massive datasets of digital pictures that have fuelled the training of many of today's facial

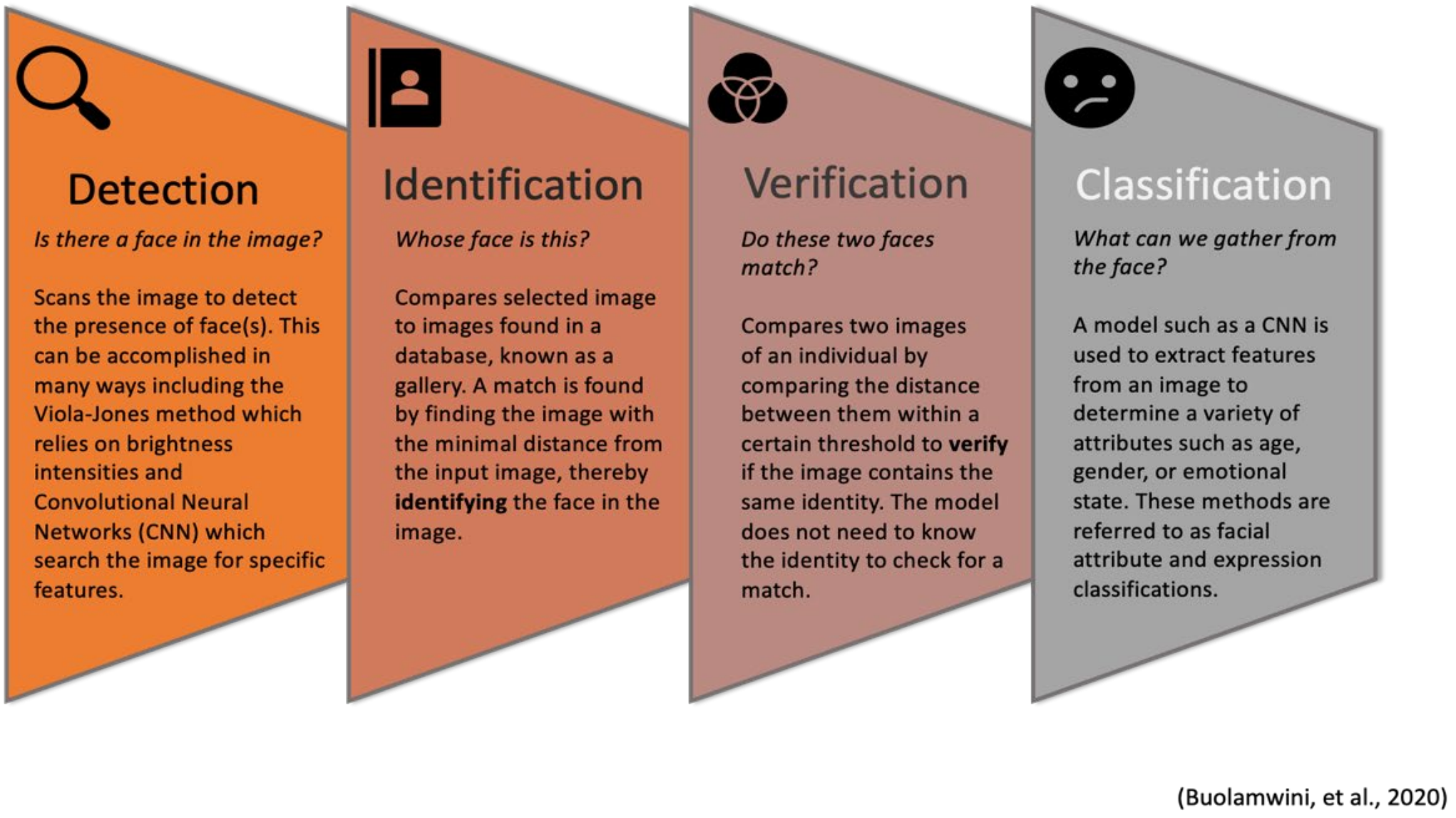

*Figure 1: Four Primary Facial Detection and Recognition Techniques*

detection and recognition models originate in this birth of the internet age.

## *A brief primer on facial analysis technologies*

But how does this all work? How do computer vision algorithms actually "see" objects in photos and images like faces or cats? Crucially, they do this in a very different way than humans do. When humans look at photo of a face, the sensory stimuli of light waves that bounce off the picture hit our retinas and travel up the optic nerve to the visual cortex. There, our brains organise this commotion of perceptual stimuli by applying concepts drawn from previous knowledge to identify and link visual patterns and properties. Our ability to connect, group and separate these elements of perception as understandable ideas enables us to form a coherent mental picture and make sense of relevant information. We are able to identify and recognise faces in a photo as a result of this collaboration between our eyes and our brains.

By contrast, when a computer vision algorithm is presented with a digital image, what it, in fact, "sees" is only a grid of pixel values (rows and columns of numbers indicating intensities of colour and brightness). In order to detect a face in a picture, a computer vision algorithm must take this two-dimensional array of integers as an input and then pinpoint those patterns in the numerical values of the pixel intensities that reliably indicate facial characteristics.

Early computer vision techniques approached this in a top-down manner. They encoded human knowledge of typical facial features and properties as manually defined rules (like "a face has two symmetrical sets of eyes and ears") and then applied these rules to the numerical grid to try to find corresponding patterns in the pixel intensity values. There was, however, a host of problems with this approach. First, translating knowledge-based rules about the human face from verbal expressions to mathematical formulae proved extremely challenging given the imprecision of natural language and the numerous ways in which different human faces could reflect the same rule. Second, this rule-based method would only really work well in uncluttered pictures in which subjects were front-facing and where environmental conditions like lighting and occlusions as well as subject-specific elements like facial orientation, position and expression were relatively uniform.

With the increasing availability of digital image data at the turn of the twenty-first century, however, top-down approaches began to give way to more bottom-up, statistics-based computer vision methods where algorithmic models were trained to learn and extract facial features and properties from large datasets. The idea here was that, given the everyday human capacity to detect and recognise faces across differing poses and under vastly varying conditions, there must be some latent features or properties that remained invariant over such differences.[2] An algorithm that could be trained on massive heaps of data to pick up or infer the most critical of these latent features, would be able to better detect and recognise human faces "in the wild" where the rigid constraints that were necessary to successfully apply handcrafted rules were infeasible.

In 2001, at the very beginning of this transition to the predominance of data-driven approaches to computer vision, researchers Paul Viola and Michael Jones created a game-changing algorithm for detecting faces, which remains widely in use even today.[3] Instead of trying to directly extract features like facial parts, skin tones, textures and fiducial points (such as the corners of the mouth or eyes), the Viola-Jones algorithm (Figure 2) scanned through digital images using very simple rectangular shapes known as Haar features—edges, lines and diagonals of different scales—to identify common patterns that allowed for rapid facial detection. Using machine learning techniques, the algorithm was trained on a large dataset of face and non-face images to narrow down the space of all possible rectangular shapes to the most important ones. These included, for instance, the horizontal edge that indicates the brightness difference between the eyes and the forehead. To detect a face, quantified, pixel value expressions of each of these features would be placed over the image's numerical grid and slid across the entire picture from box to box in order to find those areas where matching changes in brightness intensity uncovered corresponding matches with facial patterns in the image. When combined, these simple features could be applied in a cascading way to efficiently and accurately sort faces from non-faces in any given digital picture.

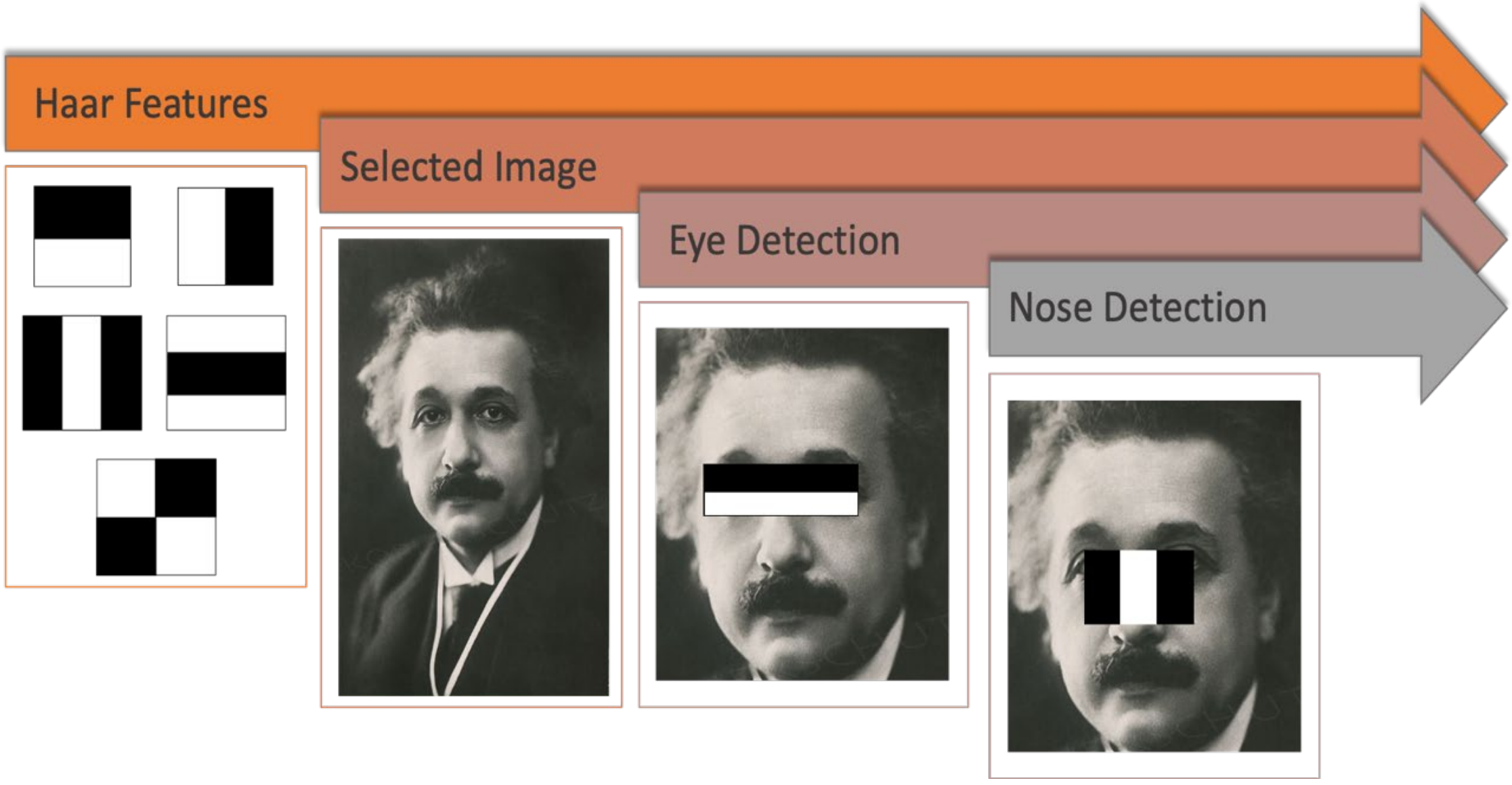

*Figure 2: Viola-Jones Facial Detection: Depiction of how Haar features are used to identify facial features such as the eyes and nose*

The Viola-Jones face detector algorithm was a bellwether in the shift of computer vision techniques from traditional, knowledge-driven and rules-based methods to data-driven machine learning approaches. Within fifteen years, a data-centred approach to facial detection and analysis called deep learning would come to dominate the scene. To grasp this transition, it is helpful to think about how these deep learning techniques moved beyond the limitations of the Viola-Jones algorithm. Whereas the latter was very good at detecting faces using repetitive scanning for simple lines and edges, it could not distinguish between faces or identify pairs of the same face among many. For this, a deeper and more nuanced set of latent facial features and properties needed to be extracted from digital images to enable comparison, contrast and analysis of the sort that would allow a computer vision model to predict whether two photos represented the same face, or whether a face in one photo matched any others in a larger pool of photos.

The algorithmic method that proved up to this task is called the convolutional neural network, or CNN for short. Similar to the Viola-Jones algorithm, CNNs break down a digital image's two-dimensional array of pixel values into smaller parts, but instead of sliding rectangular shapes across the image looking for matches, they zoom in on particular patches of the image using smaller grids of pixels values called kernels (Figure 3). These kernels create feature maps by moving step-by-step through the entire image trying to locate matches for the particular feature that the neural net has trained each of them to search for. A convolutional layer is composed of a stack of these feature maps, and any given CNN model may be many layers deep. What makes a CNN so powerful at feature extraction for facial detection and analysis is that such iterative layers are hierarchically organised. Each next layer is able to combine the features from previous layers to extract patterns that are at higher levels of complexity and abstraction. So, a first layer may cull edges, lines, dots and blobs, the next layer noses, eyes and ears, and the third fuller facial structures.

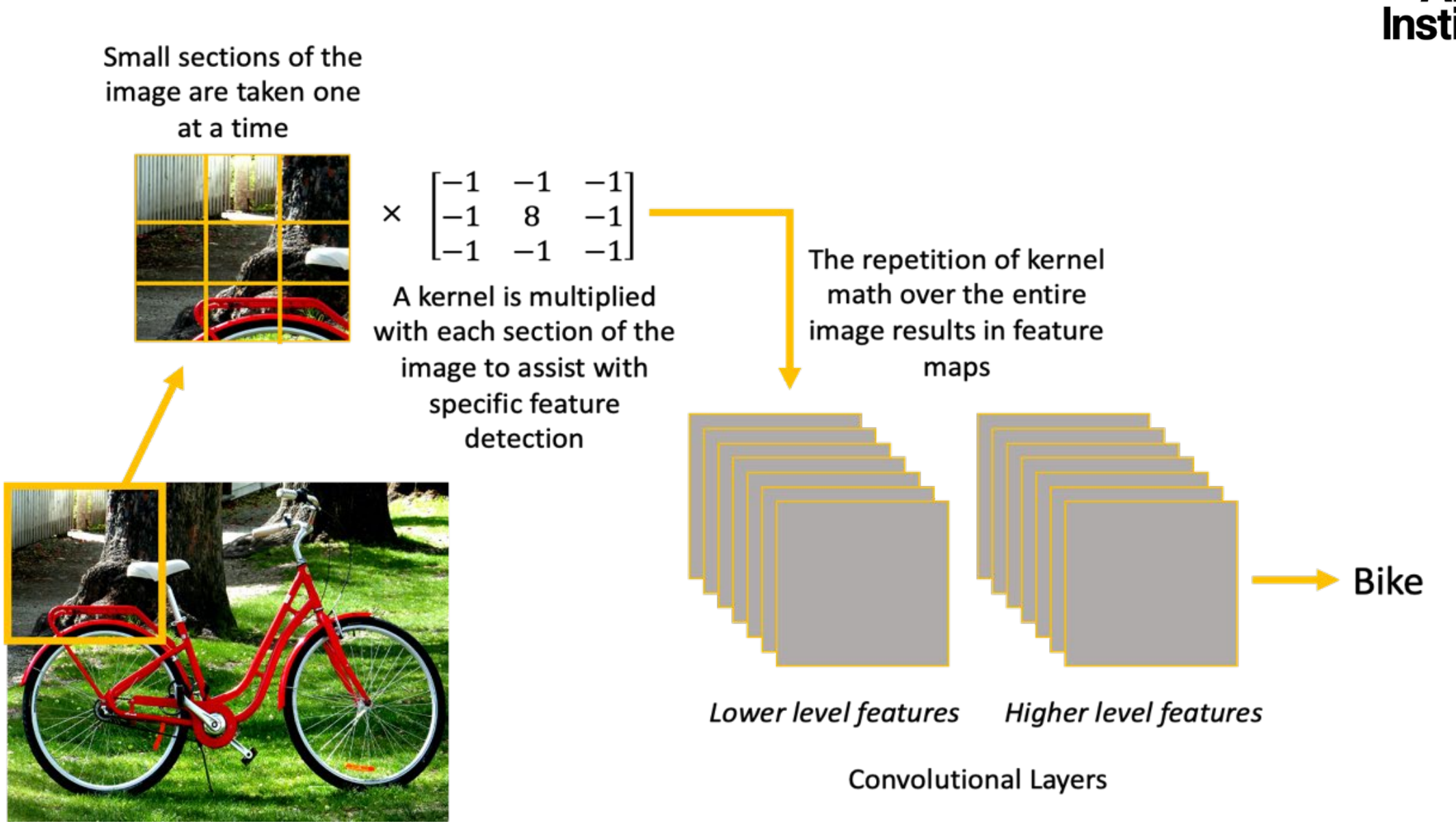

*Figure 3: Higher-level Illustration of CNN Architecture*

The key aspect of CNNs that enables data-driven feature selection and extraction is called supervised learning. Rather than starting with human-generated rules or handcrafted features, supervised learning models draw out patterns by "learning" from large datasets that contain labelled data (i.e. data that has been identified and tagged—usually by humans—as being or not being of the target class or classes that a given model is trying to predict). In facial detection and analysis models, this kind of learning occurs through the processing of vast numbers of labelled images, some containing faces and others not. When a CNN begins its training, the numerical values of its kernels are set at random (each kernel does not yet "know" what feature or pattern it is looking for). But as it processes more and more examples, the feature values of its kernels are adjusted to improve upon the mistakes it has made in each previous prediction.

This adjustment process is called backpropagation. Here, the error in the prediction—the numerical difference between the predicted answer and the actual answer of the classification problem—is used to determine how the values in each of the kernels can be nudged to minimise the error and to better fit the data. In a step-by-step procedure known as gradient descent, the algorithm goes back through all of its layers for each training example and re-tunes the pixel values of its kernels (as well as the weights of the nodes in its other decision-making layers) to optimise its correction and improve accuracy. Eventually, across thousands, hundreds of thousands, or even millions of training example adjustments, the kernels

“learn” which features and patterns to search for in out-of-sample images, and the model learns how to bring these feature representations together to make accurate predictions about what it is trying to classify.

It’s not difficult to see how the explosion of the availability of digital images on the internet fuelled the development of CNNs and other related deep learning techniques as the dominant computer vision algorithms of the big data age. Coupled with continuous gains in information processing power, the growth of large labelled datasets meant that research energies in computer vision could draw on a veritably endless reservoir of training data to optimise object detection and to identify and differentiate between individual classes. In 2005, computer scientists from Stanford and Princeton began working on ImageNet, a dataset that would become the standard-bearer of internet-scraped training data. ImageNet eventually grew to over 14 million labelled examples of over 20 thousand classes (mostly hand-labelled by crowdsourced workers from Amazon mechanical Turk), and it would provide the basis of the algorithmic model competitions that first launched the rise of CNNs in 2012.[4] By 2017, the largest of the internet-sourced datasets, Google’s JFT-300M, included 300 million images, containing over 1 billion algorithmically labelled objects drawn from 18 thousand classes. Meanwhile, in the more specialised world of FDRTs, the modest “Labelled Faces in the Wild” dataset, released in 2007 and comprised of 13 thousand images of almost 6 thousand individual identities, would be eclipsed by the likes of Oxford University’s 2018 VGGFace2 dataset, which contained over 3.3 million images of about 9 thousand individual identities (Figure 4).

## How could face detection and analysis algorithms be biased?

The story of bias in facial detection and analysis algorithms is a complicated one. It involves the convergence of complex and

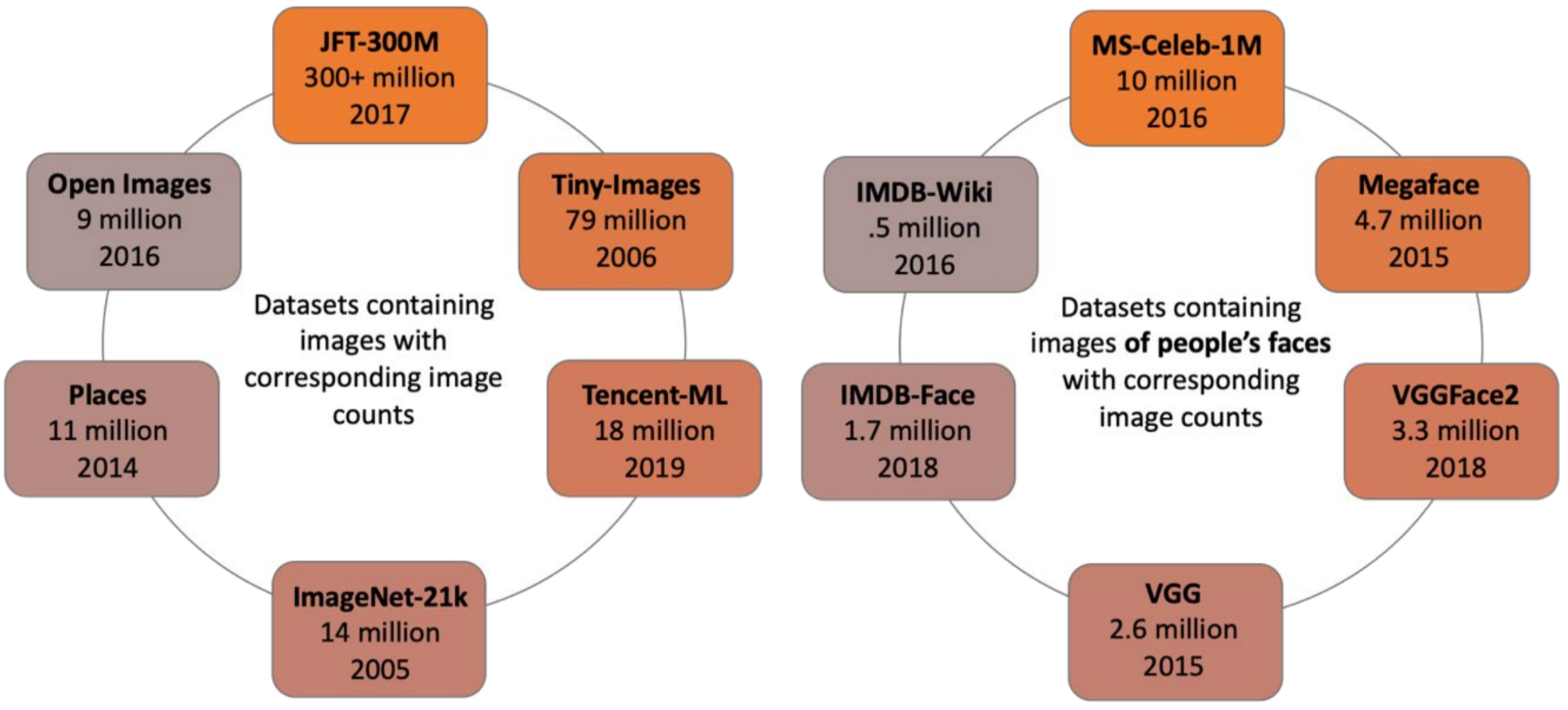

*Figure 4: Public Datasets: Size and Year of Creation Images (left) and Images with Faces (right)*

(Merler, et al., 2019; Pearson, 2019; Prabhu & Birhane, 2020)

culturally entrenched legacies of historical racism and white male privilege in visual reproduction technologies with the development of new sources of bias and discrimination arising from the novel sociotechnical contexts of digital innovation and algorithmic design and production. That is, this story of algorithmic bias involves a carry-over of certain discriminatory assumptions from the history of photography into computer vision technologies. It also involves the appearance of new conditions of discrimination that have cropped up at every stage of the evolution of FDRTs themselves—from the way that human biases have crept into model design choices to the manner in which the massive datasets that feed data-driven machine learning have tended to overrepresent dominant groups and marginalise people of colour. Let's consider these in turn.

### *Reverberating effects of historical racism and white privilege in photography*

The transition from analogue, film-based photography to digital, pixel-based photography was not a clean break in terms of the sociocultural assumptions that continued to shape the way people of different skin tones were represented in technologies of visual representation and reproduction. From its very inception, the history of photography was influenced by the privileging of whiteness as the global norm of ideal flesh tone. As early as the 1840s, portrait photographers prioritised perfecting the capture of white skin through the chemistry of film, the mechanics of camera equipment and the standardisation of lighting and procedures of film development.[5]

As scholars Richard Dyer, Lorna Roth and, more recently, Sarah Lewis, Simone Browne and Ruha Benjamin have all highlighted, considerations of ways to optimise the representation of people of colour have baldly been cast aside throughout much of the nearly two-hundred-year history of photography.[6] In the twentieth century, the privileging of whiteness in the evolution of photographic technologies led to the design of film emulsion and development techniques that were biased toward "Caucasian" skin tones. Photographs taken with colour film purchased from companies like Kodak and Fuji were insensitive to the wide range of non-white skin types and often failed to pick up significant details of the facial features and contours of those with darker skin.[7] Revealingly, up to the 1990s, Kodak's international standard for colour balance and skin-tone accuracy, the "Shirley card," used the skin colour of white female models to calibrate veritably all film developed at photo labs.

By the time Shirley cards became multi-racial and Kodak started selling its less biased Goldmax film (which it odiously claimed to be "able to photograph the details of a dark horse in lowlight"[8]), the digital photographic technologies that would soon supplant film-based photography were appearing on the scene. Digital cameras brought some immediate improvements. They increasingly allowed picture-takers to expand the dynamic range of exposure to accommodate variations in skin tone, and they enabled them to adjust and retune contrast, brightness and colour balance after photos had been taken. But, the deep-seated bias toward the default whiteness preference of visual reproduction technologies – what has been called "flesh tone imperialism"[9] – persisted. Writing in 2014, African-American photographer Syreeta McFadden emphasised this continuance of the default privileging of paleness:

> Even today, in low light, the sensors search for something that is lightly colored or light skinned before the shutter is released. Focus it on a dark spot, and the camera is inactive. It only knows how to calibrate itself against lightness to define the image.[10]

McFadden's observation reflects the fact that residua of the bias towards optimising the capture of white skin (what Browne elsewhere calls the dominant cultural "logic of prototypical whiteness"[11] and Benjamin simply "the hegemony of Whiteness"[12]) have survived into the era of smart cameras and facial recognition algorithms. As digital cameras began to incorporate FDRTs, problems of the preset prioritisation of whiteness cropped up in kind. In 2009, an Asian American blogger protested that the Nikon Coolpix S360 smart camera she had bought for her mother kept registering that members of her family were blinking, while the photos clearly showed that all eyes were wide open.[13] What likely lay behind the repeated error was a facial detection algorithm that had not been sufficiently designed to discern the few-pixel difference between narrow shaped eyes and closed ones—either the inadequately refined pixel scale of the feature detection tool or downsampling (a coarsening of the picture array) meant that Asian eye types would often go unrecognised.[14] That same year, workers from a camping shop demonstrated in a viral YouTube video that a face-tracking webcam from a Hewlett Packard laptop easily followed the movements of the lighter-skinned employee's face but went entirely still when the darker-skinned face replaced it.[15] This led the man whose face went invisible to the device, Desi Cryer, to object: "I'm going on record and saying Hewlett-Packard computers are racist..."[16]

HP's response to Cryer's claims is illustrative of the lingering forces of racial bias that have continued to influence the design of facial detection and analysis algorithms. After publicly acknowledging the problem, HP officials explained:

> The technology we use is built on standard algorithms that measure

> the difference in intensity of contrast between the eyes and the upper cheek and nose. We believe that the camera might have difficulty "seeing" contrast in conditions where there is insufficient foreground lighting.[17]

There are two kinds of algorithmic bias that arose in HP's action of releasing its Mediasmart webcam—both of which have their origins in the cultural logic of prototypical whiteness and its corollary privileging of a light-skinned reality. First, the designers of the algorithm chose a feature detection technique that clearly presupposed normal lighting conditions optimal for "seeing" the contrastive properties of lighter-skinned faces. Rather than starting from the priority of performance equity (where the technical challenges of dynamic exposure range, contrast and lightening are tackled with equal regard for all affected groups), the designers focused on marshalling their technical knowledge to deliver maximal results for those whose pale, light-reflecting faces were best recognised using differentials in intensity contrast.[18] Second, the starting point in a cultural hegemony of pervasive whiteness, created a significant apathy bias where the technical challenges of securing equitable performance were neither prioritised nor proactively pursued. HP technologists neglected to reflectively anticipate the risks of harm to various impacted communities in cases of algorithmic failure. Likewise, they appeared not to have thoroughly tested the technology's differential performance on non-white groups to safeguard against discriminatory outcomes. They just released the system come what may.

## *New forms of bias in data-driven computer vision*

The missteps that led to HP's production of a discriminatory facial detection algorithm clearly illustrate the way biases in design choices and motivations can end up perpetuating underlying patterns of racism. These lapses, however, do not even begin to scratch the surface of the novel forms of discrimination that have attended the transition to data-driven computer vision in the era of big data. On the face of it, the shift from rule-based methods and handcrafted features to data-centric statistical techniques held the promise of mitigating or even correcting many of the ways that biased human decision-making and definition-setting could enter into the construction of algorithmic models. Troublingy, however, this has not been the case. The coming of the age of deep learning and other related ML methods in facial detection and analysis has been typified both by the cascading influences of systemic racism and by new forms of discrimination arising from unbalanced data sampling, collection and labelling practices, skewed datasets and biased data pre-processing and modelling approaches.

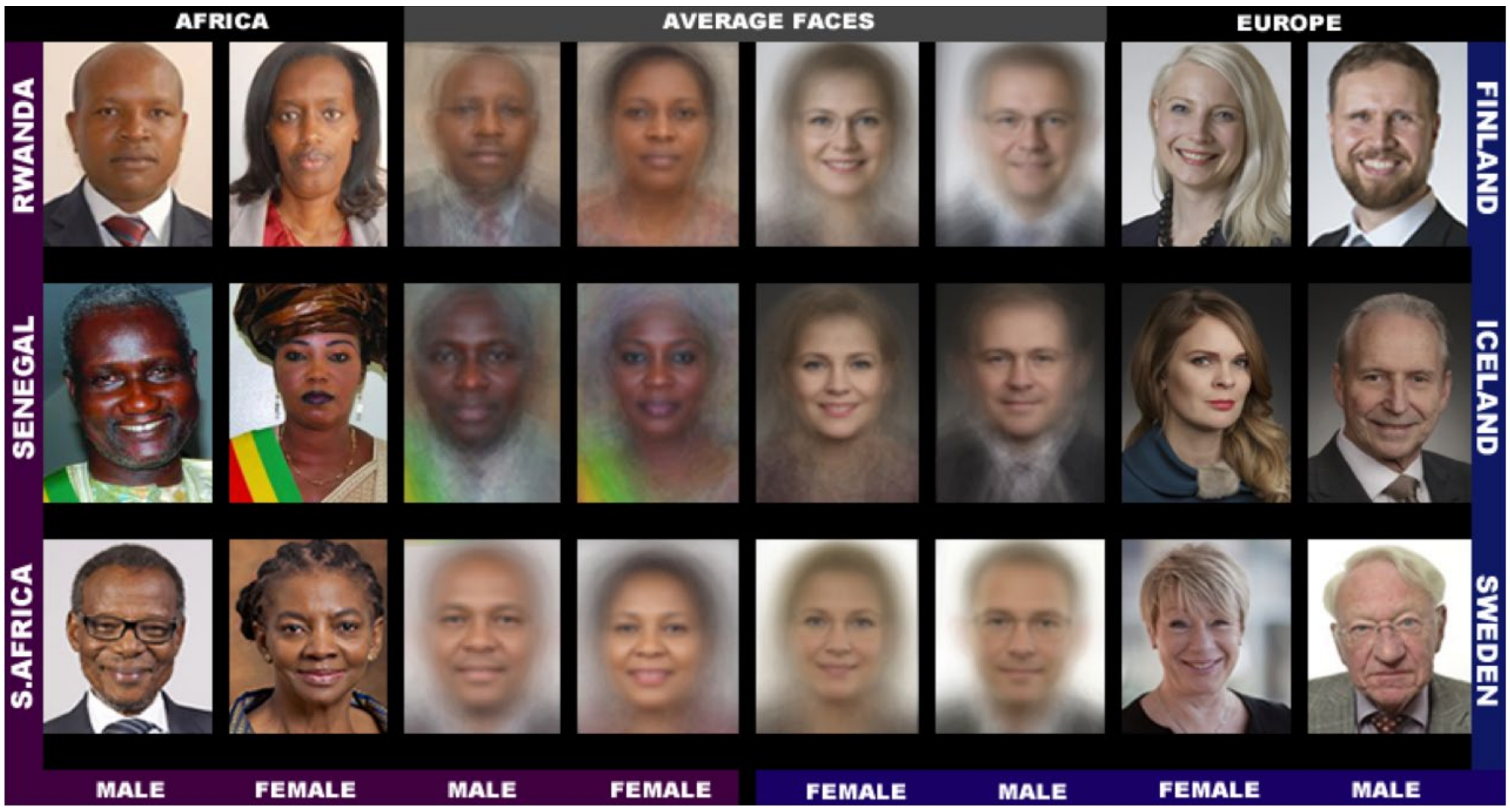

*Figure 5: Pilot Parliaments Benchmark (PPB) Dataset* (Buolamwini & Gebru, 2018)

From the earliest moments of the commercial development of data-driven computer vision and facial analysis models, researchers and watchdogs have pointed out that these systems tend to perform differently for different demographic groups. A series of studies put out by the National Institute of Standards and Technology in the US from 2002 to 2019 demonstrated significant racial and gender biases in widely used facial recognition algorithms.[19] Many others have similarly shown dramatic accuracy differentials in these kinds of applications between gender, age, and racial groups—with historically marginalised and non-dominant subpopulations suffering from the highest levels of misidentification and the greatest performance drops.[20] Building on this research in an important 2018 paper, Joy Buolamwini and Timnit Gebru highlighted the compounding effects of intersectional discrimination (i.e. discrimination where protected characteristics like race and gender overlap) in facial analysis technologies trained on largescale datasets.[21] Buolamwini and Gebru audited commercial gender classification systems produced by Microsoft, IBM, and Megvii's Face++ for performance disparities and showed that the rate of misclassification for darker-skinned woman was, at points, thirty-five times or more higher than for white men.

The method that Buolamwini and Gebru used to expose the algorithmic bias in these systems also shed light on the main sources of discrimination. To test the performance of each respective application, they created a benchmark dataset that was composed of balanced subgroups of genders and skin types (Figure 5). This helped to lay bare that, while claims of impressive overall performance seemed to indicate that companies like IBM and Microsoft had produced extremely accurate facial analysis algorithms, when the outcomes were broken down by subgroups, some alarming disparities emerged. For instance, Microsoft's FaceDetect model demonstrated an overall error rate of 6.3% on its gender classification tasks. However, when its performance was analysed in terms of the intersection of gender and race, the results showed that whereas the application had a 0% error rate for light-skinned males, it had a 20.8% error rate for dark-skinned females. Not only did this reveal evidence that the largescale datasets used to train and benchmark these systems were underrepresenting people of colour and women, it uncovered the widespread lack of attention being paid by model designers to the measurement of performance disparities for historically marginalised groups.

Representational biases in largescale vision datasets arise, in part, when the sampled images that comprise them are not balanced enough to secure comparable performance for members of the various demographic groups that make up affected populations. The most widely used of these datasets have historically come from internet-scraping at scale rather than deliberate, bias-aware methods of data collection. For this reason, the heaps of training material that have been fed into the engines of big-data-driven facial analysis in its first decade of life have failed to reflect balanced representations of the diverse populations that such technologies impact. Rather, they have largely reflected the power relations, social hierarchies and differential structures of privilege that have together constituted the sociocultural reality from which those data were extracted in the first place. A recent study by IBM researchers on facial diversity in datasets, in fact, shows that out of the eight most prominent, publicly available largescale face image datasets, six have more male images than female ones and six are composed of over eighty percent light-skinned faces.[22] The skewing of these datasets toward "pale male" visages is just another example of how legacies of structural racism and sexism have kept an active grip on technologies of representation in the age of big data.

This grip, however, is not limited to the social and historical conditioning of data. It also manifests in the motivations and intents of the designers and developers behind the machine learning systems that are trained on biased datasets. One of the central problems that has enabled systemic inequities to creep into these datasets seemingly unchecked is the way that neither dataset builders nor algorithm designers prioritised the attainment of technical capacities to identify, understand and redress potentially discriminatory imbalances in the representation of the demographic and phenotypic characteristics of their data. For instance, as of last year, out of the ten biggest largescale face image datasets, none had been labelled or annotated for skin colour/type, rendering performance disparities across different racial groups all but invisible to those who used these datasets without further ado to train, test and benchmark their algorithmic models.[23]

These biases of apathy and omission derive largely from the complacency of technology producers who occupy a privileged station as members of the dominant group and hence do not face the adverse effects of discriminatory differential outcomes. Such motivational deficiencies have turned all-the-more insidious as the societal impacts of the widening use of these technologies have increased. To take two examples, performance disparities in commercial grade facial recognition systems that disproportionately generate false-positives for people of colour can now lead to unjust arrests and prosecutions. Likewise, differential accuracy in the outputs of the computer vision component of autonomous vehicles can now kill dark-skinned pedestrians.[24]

However, beyond the biases that emerge from the lack of motivation to label and annotate largescale face image datasets in a fairness-aware manner (and the corollary failure to proactively correct accuracy differentials that may harm marginalised groups), issues of prejudice and misjudgement surrounding *how* these data are labelled and annotated can also lead to discrimination. Human choices about how to classify, breakdown and aggregate demographic and phenotypic traits can be intensely political and fraught endeavours.[25] For instance, the categories of "race," "ethnicity," and "gender" are highly contestable, culturally-laden and semantically unstable. Simple binary categorisations of "race" and "gender" (like white/non-white and male/female) may impose the cultural norms of prevailing pale-skinned, cis-gender groups.

More fundamentally, the encoding of ascribed human traits for the purpose of the physiological classification of the human body has long been a pseudo-scientific method of discriminatory and racialising social control and administration.[26] In facial recognition technologies, such an imposition of racial categories leads to what Browne has termed "digital epidermalization"—a kind of epistemic enactment of subcutaneous scientific racism.[27]

In their analysis of the 20 thousand image UTKFace dataset published in 2017, Kate Crawford and Trevor Paglen illustrate some of these difficulties. In the UTKFace dataset, race is treated as falling under four categories: White, Black, Asian, Indian, or "Others"—a classificatory schema that, as Crawford and Paglen point out, is reminiscent of some of the more troubling and problematic racial taxonomies of the twentieth century such as that of the South African Apartheid regime. Similarly precarious is the way that the UTKFace dataset breaks down gender into the 0(male)/1(female) binary, assuming that the photographic appearance of an either/or-male/female self is an adequate marker for labellers to ascribe gender to faces in images, thereby excluding by classificatory fiat the identity claims of an entire community of non-binary identifying people.[28]

All of this speaks to the fact that justifiably deciding on the categories, groupings and gradients of demographics like race, ethnicity, and gender for labelling and annotation purposes is a difficult and discursive process that requires the continuous participation of affected communities, debate and revision. Confronting these issues constructively involves airing out the often-secreted historical patterns of discrimination and the underlying dynamics of power that infuse identity politics and struggles for recognition. It also involves stewarding an open, respectful and interactive process where reasons and normative justifications can be offered for classificatory choices made. Buolamwini and Gebru, in this latter respect, dedicate an entire section of their paper on intersectional bias to offering their rationale for choosing the Fitzpatrick/dermatological classification of skin types for their benchmark dataset over the kind of demographic classificatory framework used in the UTKFace dataset. The deeper implication of this general need for more deliberative and discursive approaches to taxonomising, classifying and labelling face image datasets is that bias-aware practices demand an active acknowledgement of the open, constructed and social character of data. They demand a continuous and collective negotiation of data ontologies, categories, properties and meanings through inclusive interaction, equitable participation and democratic dialogue.

## Wider ethical issues surrounding the design and use of facial detection and analysis systems

Much of the discussion around the ethical issues that have arisen with the emergence of data-driven biometrics and facial analysis technologies has centred on the real-world harms they cause and how these might be remedied (Figure 6). Such a conversation has largely taken the existence of FDRTs for granted as a necessary means for securing public safety or improving efficiency and focussed instead on how to right the things that are going wrong with present practices of designing and deploying these technologies. Still, it is important to recognise that there is an even more basic set of ethical questions that are just as, if not more, vital to consider.

These questions surround the very justifiability of the development and use of FDRTs in the first instance. This more rudimentary line of inquiry raises questions like: Should we be doing this to begin with?

Are these technologies ethically permissible to pursue given the potential individual and societal impacts of their broad-scale development? Do technologists and innovators in this space stand on solid moral ground in flooding society with these expanding capacities to globally automate unbounded personal identification and ubiquitous smart surveillance? Is the programmatic endeavour to create systems that dubiously claim to infer human emotions and personality characteristics from facial properties, shapes and expressions even justifiable?

Ideally, deliberations concerning this second set of questions about ethical justifiability should be undertaken in advance of the development of the technologies themselves. Such an *ex ante* approach should involve getting out ahead of their real-world use and thinking in an anticipatory way about whether or not they should be pursued in view of the range of risks they may pose to affected individuals and communities. Troublingly, however, the current state of play in the innovation ecosystem of algorithmic facial analysis almost dictates that one has to think about issues of ethical justifiability retrospectively rather than prospectively—that is, by looking immediately backwards at the noxious effects of already existing innovation practices.

The recent history of the explosive growth of the industry that builds facial processing systems and markets them to law enforcement and intelligence agencies, the military and private corporations demonstrates that the "move fast and break things" attitude first explicitly championed by Facebook in 2014 has operated for facial

| | |
|---|---|
| Public Interest | Is the technology being deployed in the public interest? |
| Effectiveness | Is the technology justified as an effective tool for its intended purpose of identification? |
| Avoidance of Bias and Algorithmic Injustice | Technology should not exhibit undue bias, i.e. discriminatory outcomes or treatment; misrecognition. |
| Impartiality and Deployment | Technology should never disproportionately target certain events without "compelling justification." |
| Necessity | Technology should not infringe on individuals' rights through monitoring or scrutinisation. |
| Proportionality | Technology can only be permissible if the benefits are proportional to losses of liberty and privacy. |
| Impartiality, Accountability, Oversight and the Construction of Watchlists | Watchlists should be subject to oversight by an independent body along with being free from bias. |
| Public Trust | Those involved with the technology should participate in public consultations and provide rationale. |
| Cost-effectiveness | Could the resources required for the technology be put to better use elsewhere? |

(BFEG Facial Recognition Working Group, 2019)

*Figure 6: BFEG's Ethical Principles for Live Facial Recognition*

analysis vendors and developers as the rule rather than the exception. For instance, recent reports about the facial recognition start-up Clearview AI have exposed that the company has sold its capacity to rapidly identify individuals by sifting through a proprietary database of three billion face images—many of which it has unlawfully scraped from social media websites and applications—to hundreds of police forces, the US Department of Homeland Security, the US Federal Bureau of Investigation, and numerous government agencies and retail companies across dozens of countries.[29] What is more, it has done all this with no transparency, no regulatory oversight, no public auditing for discriminatory performance disparities, no checks on violations of consent or privacy protections and no substantive accountability. Examples like this seem to be multiplying, at present, as blinkered and reckless innovation practices like those of Clearview AI couple with the corresponding embrace by equally feckless government agencies and corporations of ethically and legally suspect facial analysis products and services (relevant studies are listed in Table 1).[30] The gathering momentum of this ostensibly inevitable race to the ethical bottom would appear to indicate an enfeebling if not disqualification of *ex ante* approaches to technology governance that seek to anticipate possible harms in order to critically reflect on their first-order permissibility. A sneaking sense of technological determinism—the reductive intuition that technology of itself holds the leading strings of society—is difficult to avoid in the current climate. This often leads to the fatalistic feeling that once a technological innovation like automated facial recognition appears it is, for better or worse, here to stay.

*Table 1: Major Public Reports on Facial Recognition Technologies*

| Title | Author(s) | Year | Contribution | Link |
|---|---|---|---|---|
| The Perpetual Line-Up: Unregulated Police Face Recognition in America. | Clare Garvie, Alvaro Bedoya, and Jonathan Frankle<br><br>Georgetown Center on Privacy and Technology | 2016 | A comprehensive survey of law enforcement, it demonstrates that unregulated facial recognition technologies affect over 117 million American adults. Additionally, the FBI biometric database is found to include law-abiding citizens. The authors explain risks of FRTs to rights and liberties and calls for regulation of these technologies – Model Face Recognition Act. | https://www.perpetuallineup.org |
| Face Off: the lawless growth of facial recognition in UK policing | Big Brother Watch | 2018 | Provides explanation of the use of FRTs by UK police departments, how the technologies function, and the threat to legal and human rights that the increased use of these technologies poses. Report also demonstrates how FRTs disproportionately misidentify minorities and women, and 95% of police's 'matches' have misidentified individuals. | https://bigbrotherwatch.org.uk/wp-content/uploads/2018/05/Face-Off-final-digital-1.pdf |
| Interim Report on Live Facial Recognition | London Policing Ethics Panel | 2018 | Reports on the status of trials by police forces using Live Facial Recognition (LFR), along with highlighting ethical issues that are necessary to consider before wider adoption of this technology. The Ethics Panel draws attention to several topics including engagement of the public, inaccurate identification, creation of a watchlist, overt and covert surveillance and limited trials leading to 'unlimited adoption.' | http://www.policingethicspanel.london/uploads/4/4/0/7/44076193/lpep_report_-_live_facial_recognition.pdf |
| An Evaluation of South Wales Police's Use of Automated Facial Recognition | Bethan Davies, Martin Innes, and Andrew Dawson | 2018 | Documents policing outcomes attributable to AFR in identifying suspects, examines organisational, system, and operator performance of the technology, and explains the ethical and legal issues associated with AFR deployments. Report concludes that technology should be renamed Assisted Facial Recognition as opposed to Automated. | https://static1.squarespace.com/static/51b06364e4b02de2f57fd72e/t/5bfd4fbc21c67c2cdd692fa8/1543327693640/AFR+Report+%5BDigital%5D.pdf |
| Garbage in, Garbage Out: Face Recognition on Flawed Data | Claire Garvie<br><br>Georgetown Center on Privacy and Technology | 2019 | This report highlights questionable law enforcement methods of inputting doctored or altered images and forensic sketches into facial recognition systems. It demonstrates how these misuses and abuses of the technology expose reckless practices of using poor quality and tampered-with data in high impact contexts where stringent rules about the procedures necessary for accurate biometric matching should be in place. | https://www.flawedfacedata.com |
| Independent Report on the London Metropolitan Police Service's Trial of Live Facial Recognition Technology | Pete Fussey and Dr. Daragh Murray<br><br>University of Essex Human Rights Centre | 2019 | Extensive report on test deployments of LFR conducted by the Metropolitan Police Service including details about governance, practices and procedures, and human rights compliance. The report calls attention to the MPS' claim of legal authorisation for the use of LFR along with the lack of publicly available guidance on the use of LFR. Both contribute to an overall need to reform how new technologies are trialled and introduced. | http://repository.essex.ac.uk/24946/1/London-Met-Police-Trial-of-Facial-Recognition-Tech-Report-2.pdf |
| Facial Recognition Technologies in the Wild: A Call for a Federal Office | Erik Learned-Miller, Vicente Ordóñez, Jamie Morgenstern, and Joy Buolamwini<br><br>Algorithmic Justice League | 2020 | Gives explanation of how current legislation does not cover the entire scope of FRTs. The authors call for a new federal office that is created to investigate the complex trade-offs of FRTs, draft legislation, ensure adherence to ethical standards, and provide coordination. | https://global-uploads.webflow.com/5e027ca188c99e3515b404b7/5ed1145952bc185203f3d009_FRTsFederalOfficeMay2020.pdf |
| Facial Recognition Technologies: A Primer | Joy Buolamwini, Vicente Ordóñez, Jamie Morgenstern, and Erik Learned-Miller<br><br>Algorithmic Justice League | 2020 | Provides definitions of FRTs specifically catered to a non-technical audience to increase awareness and understanding of these technologies. Basic definitions and examples are included in this primer, along with use cases and descriptions of fundamental technical concepts and challenges. | https://global-uploads.webflow.com/5e027ca188c99e3515b404b7/5ed1002058516c11edc66a14_FRTsPrimerMay2020.pdf |

In reality, though, things are playing out a little differently. As recent waves of corporate back-peddling, Big Tech moratoria, successful litigations and local facial recognition bans have shown, the critical force of retrospective anticipation that responds to societal harms by revisiting the legitimacy and justifiability of facial analysis technologies is playing a major role in reining them in. In US cities from Oakland, San Francisco, and Berkeley, California to Cambridge, Brookline and Somerville, Massachusetts, the use of facial recognition systems by law enforcement agencies has been prohibited,[31] while, in Portland, Oregon, the city council has gone a significant step further, prohibiting the use of facial recognition by both local authorities and private corporations.[32]

In June 2019, Axon, a major producer of police body cameras, responded to a review of its independent ethics board by banning the use of facial analysis algorithms in its systems.[33] Similarly, following the killing of George Floyd and acknowledging the link between facial analysis technologies and legacies of racism and structural discrimination, Microsoft and Amazon announced moratoria on their production of facial recognition software and services,[34] and IBM also announced that it is getting out of the business entirely.[35] Even more recently, the Court of Appeal in South Wales ruled that the local police force could no longer use its automated facial recognition system, AFR Locate, on the grounds that it was not in compliance with significant aspects of the European Convention on Human Rights, data protection law, and the Public Sector Equality Duty.[36] Taken together, these various high-profile course corrections intimate that society is still capable of pumping the brakes in order to reclaim its rightful place at the controls oftechnological change. Though not optimal, the capacity to challenge the ethical justifiability of facial analysis technologies through retrospective anticipation, might as yet be a powerful tool for spurring more responsible approaches to the rectification, reform and governance of these technologies. When community-led, deliberatively-based and democratically-shepherded, such critical processes can also be a constructive force that restores a degree and manner of participatory agency in socio-technical practices that have gone awry. It becomes critical, in this respect, to gain a clearer view of the distinct ethical issues that are surfacing amidst the growing resistance to current delinquencies and misdeeds in the complex ecology of the facial recognition technology supply chain. Not only do these issues form the basis of claims being made against the felt moral injuries suffered in the present, they are, in no small measure, steering the course adjustments that may define more progressive and responsible approaches to the development of facial analysis innovation in the future. Let's move on then by spelling out some of the more significant of these issues, though remaining mostly focussed on the dimensions of bias and discrimination.

### *Double-barrelled discrimination*

The primary concentration of this explainer has been on unpacking how problems of bias and discrimination have come to manifest in computer vision and facial analysis systems as these have transitioned into an epoch dominated by data-driven machine learning. However, an understanding of the way these mechanisms of bias and discrimination work within the range of facial analysis technologies does not yet flesh out the nature of the moral harms suffered by those whose lives are directly impacted by their misuse and abuse. To address this latter problem, we need to better grasp the character of ethical claims that those who are adversely affected by these misuses and abuses can make. There are, at least, two kinds of such claims. First, there are claims for distributive justice— claims that turn on the demand for equal moral regard in the allocation of the benefits and risks emerging from the use of a given technology. And, second, there are claims for recognitional justice—claims that turn on the demand for reciprocal respect in the acknowledgment of the equal moral worth of individual and group identity claims.[37] Along these lines, distributive injustices occur when members of a subordinated or discriminated-against demographic or social group are refused access to benefits, resources or opportunities on the basis of their affiliation with that group (Figure 7). These have appropriately been called "harms of allocation."[38] Recognitional injustices, by contrast, happen when the identity claims of members of a subordinated or discriminated-against demographic or social group are denied or violated in ways that reaffirm and augment their marginalised position. These have been called "harms of representation."[39]

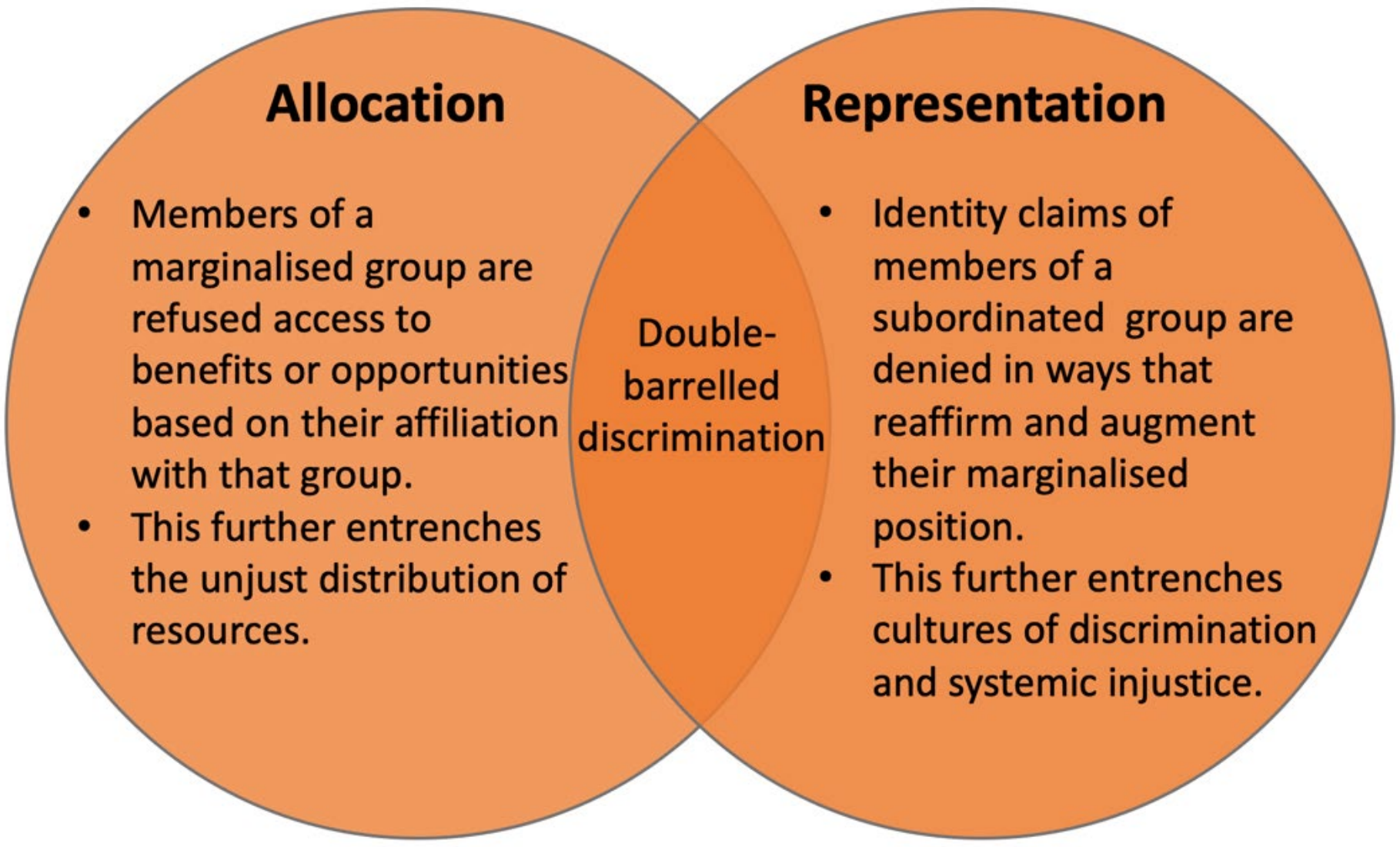

*Figure 7: Comparison of Harms of Allocation vs. Representation*

The history of FDRTs is rife with both sorts of harms, but, more often than not, they come together as a kind of double-barrelled discrimination. For instance, in the case of Desi Cryer and the discriminatory HP Mediasmart webcam, the failure of the technology to detect his face, while at the same time easily picking up and tracking the face of his white co-worker, rendered him invisible on the basis of the colour of his skin, a blatant representational harm that led him to justifiably assert that HP computers were racist. But in addition to this recognitional injustice, Cryer was also denied the benefit of the technology's intended utility, namely, of enriching interpersonal digital communication by following the movements of users' faces. The coalescence of allocational and representational harms here was multiplicative: he was refused the webcam's benefit in virtue of his discriminatory exclusion from the very dimension of enhanced social connection for which it was built.

Another, more recent, example of this kind of double-barrelled discrimination has arisen in the UK Home Office's online passport photo checking service. First launched in 2016, the system uses an algorithm to scan uploaded face images and assess whether or not they are of sufficient quality to be used in a person's passport.[40] However, in several very public instances that have been shared widely on social media and covered by the press, the facial detection and analysis component of the technology has failed to successfully process the photos of passport applicants who have darker skin-types. It inaccurately returned error message for them like "it looks like your mouth is open" and "it looks like your eyes are closed" when there was evidently nothing wrong with the uploaded images. In one case, occurring this past February, Elaine Owusu received the former error message with the additional prompt: "you can still submit your photo if you believe it should be accepted (for example, if there is a medical reason why your mouth appears to be open).” When she shared this message on social media, it triggered a slew of outraged responses such as, “A medical reason...Yh, I was born black,” “violation at the highest level,” and “this is a form of systemic racism and it’s incredibly sad to see...”.[41]

While these reactions signal warranted indignation in the face of the representational harm done to Owusu, they do not touch upon the second valence of distributive injustice that is also clearly at play here. When an automated system that is built with the intended purpose of making an administrative process more efficient and adding a convenience to a burdensome process systematically achieves its goals for some privileged social group or groups but does quite opposite for marginalised ones (i.e. increases the burden of time and effort needed to complete the same process), it metes out discriminatory harm of an allocational kind. Referring to a similar outcome that occurred several months earlier, when the Home Office’s passport photo checking algorithm failed to work for another darker-skinned user, a representative of the UK’s Equality and Human Rights Commission commented: “A person’s race should not be a barrier to using technology for essential public services.”[42]

Both the Cryer and Owusu examples well illustrate the two-pronged character of the moral harm that can be inflicted by the use of biased and discriminatory FDRTs. Even so, they do not yet plumb the depths of the damage that can be done by this type of double-barrelled discrimination in the lived experience of those adversely impacted. Take, for instance, the wrongful arrest and imprisonment for larceny of Robert Julian-Borchak Williams in Detroit, Michigan earlier this year after the police had blindly followed the lead of a faulty facial recognition system.[43] Williams, an African-American male, was accosted by officers on his front lawn in the presence of his wife and two young daughters, following the erroneous match of a picture of his held within the state’s facial recognition database with a grainy surveillance video from a boutique that had been robbed over a year earlier. Once at the detention centre, Williams, who had not even been questioned about whether he had an alibi before the arrest, was shown a series of image pairs that supposedly linked him to the crime. He held the surveillance camera pictures up to his face and pointed out that they looked nothing like him: “This is not me...You think all black men look alike?”[44] The detectives, on Williams’ account, agreed that the photos did not match, conceding that “the computer got it wrong.” But, he was not immediately released. Instead, he was kept in custody for thirty hours, compelled to post a personal bond to be let out and faced an arraignment date in the Wayne County court.[45] Though the prosecutors in Williams’ case eventually dropped the charges, the representational and allocational harms had been done.

Williams felt stigmatised. His boss insisted that he not tell his co-workers about the incident. He told a reporter: “My mother doesn’t know about it. It’s not something I’m proud of...It’s humiliating.”[46] But the trauma of the arrest in front of Williams’ home, amidst his wife and young children and in plain view of neighbours, raises a real possibility of devastating long-term impacts on him and his family. Already, shortly after the incident, one of his daughters started pretending that she was a police officer who was arresting her father for theft and incarcerating him in the living room of their family home. Aside from these crushing psychic and emotional aftereffects, Williams paid a big price in terms of the burden of time, effort, and resource needed to clear his name in wake of the Detroit Police Department’s inept overreliance on an unsound and biased facial recognition technology.

### *Gateway attitudes*

The brazen attitude of the Detroit detectives, who arrested and detained Williams based on an automated facial recognition match and very little else, flags up a separate but related ethical issue. This surrounds the complacent approach of organisations, agencies and companies who are actively procuring and using FDRTs while simultaneously failing to address or rectify known hazards of discrimination, bias and differential accuracy that may lead these systems to disparately harm vulnerable and marginalised communities. Such motivational biases of those in privileged positions, who assume an apathetic or evasive stance on remediation, could allow for these kinds of delinquencies and abuses in the deployment of facial analysis technology to continue to spread frictionlessly and into ever more impactful application areas.

By the time Robert Williams was arrested in January of 2020, the many problems associated with racial bias and performance disparities in automated facial recognition systems had become well known to the Detroit Police Department and to its vendors. In 2017, when steps were taken to integrate facial recognition technologies into the Detroit’s Project Green Light video crime monitoring program, civil liberties organisations like the ACLU expressed their concerns about the tendencies of these technologies to disproportionately generate false positives for minority groups—potentially leading to an increase in wrongful arrests and prosecutions.[47] Assistant Police Chief James White responded, at the time, that software would not be mishandled but rather applied as, “an investigatory tool that will be used solely to investigate violent crimes, to get violent criminals off the street.”[48]

However, by mid-2019, Detroit detectives had proven this wrong. On July 31 of that year, police surrounded the car of twenty-five-year-old Michael Oliver, put him in handcuffs and incarcerated him for two-and-half days.[49] Oliver's face had been identified as a match when a Detroit detective investigating a felony larceny had an image from a cell phone video involved in the incident run through the state's facial recognition database. Despite the fact that Oliver's facial shape, skin tone, and tattoos were inconsistent with the pictures of the real suspect, he was put in a digital line-up, identified by a witness on that basis, and arrested without further investigation. A court document from a lawsuit filed by Oliver contends that, once he was incarcerated, detectives "never made any attempt to take a statement from [him] or do anything to offer [him] a chance to prove his innocence."[50] Detroit prosecutors eventually dropped the case. Williams' later arrest was part of a pattern.

More precisely, it was part of a systemic pattern of derelict behaviour rooted in the apathetic tolerance of discriminatory harm. The blatant violations of the rights and civil liberties of Oliver and Williams by a police force, whose leadership was well aware of the pitfalls of inequity deriving from their facial recognition software betokens an unacceptable complacency in the willingness of public authorities to proceed with using technologies that they know run rough shod over cherished freedoms, identity-based rights and civic protections. A similarly complacent attitude has been demonstrated by the Home Office in its launch and continued use of its passport checking service. As first reported in the *New Scientist*, a Home Office response to a freedom of information (FOI) request sent in July of last year about "the skin colour effects" of its passport checking service revealed that the Government had carried out user research, which surfaced disparate performance for ethnic minorities.[51] However, the service was rolled out anyway, because, as the Home Office's reply to the FOI request claimed, "the overall performance was judged sufficient to deploy." Reacting to this revelation, the UK's Equality and Human Rights Commission commented: "We are disappointed that the government is proceeding with the implementation of this technology despite evidence that it is more difficult for some people to use it based on the colour of their skin."[52]

The bigger concern that emerges here has to do with what might be thought of as a gateway attitude: a complacent disregard for the gravity of discriminatory harm that enables organisations, agencies, and corporations in stations of power to pursue, procure and use technologies like facial recognition systems that are liable to subject marginalised and subordinated groups to distributional and recognitional injustices without sanction. This is especially troubling in the case of the Home Office where patterns of past behaviour suggestive of this sort of disregard are converging with ambitions to radically widen the scope of government use of facial recognition applications. The Home Office has recently withdrawn its visa application screening tool after digital and human rights advocates pointed out its years-long tendency to discriminate against suspect nationalities and ethnic minorities. This is of a piece with the retreat of the South Wales police force from using live facial recognition technologies (a project underwritten and funded by the Home Office) in the wake of the Appeal Court ruling that it failed to comply with data protection law, human rights conventions, and the Equality Act. In both of these instances, and in the case of the passport checking service as well, the Home Office has not merely demonstrated insensitivities to structural racism, systemic discrimination and algorithmic bias, it has actively taken forward digital innovation projects with a disturbing lack of attention paid to the anticipatory preclusion of potential violations of citizens' rights and freedoms, to the adverse ethical impacts that algorithmic applications may have on affected individuals and communities, and to the proactive rectification of known risks.

What makes these indicators of a complacent disregard for the gravity of discriminatory harm all-the-more worrisome is the potential that they enable future expansionism. Illustratively, the Home Office is currently an official project partner of a five-year research programme, FACER2VM, that aims to "develop unconstrained face recognition" for a broad spectrum of applications and aspires to soon make the technology "ubiquitous."[53] The problematic track record of the Home Office on pre-emptively safeguarding civil liberties and fundamental rights together with concerns already expressed by the UK's Biometrics Commissioner in 2018 and, more recently by its Surveillance Camera Commissioner about the department's lax attitude to setting up legislative guardrails,[54] should give us significant pause.

## *Wider ethical questions surrounding use justifiability and pervasive surveillance*

The complacent mind-set of technologists, researchers and corporate and government officials toward discriminatory harm has, in no small measure, allowed them to converge as social catalysts of the ever more pervasive uptake of facial analysis applications. Nevertheless, the expanding adoption of these technologies has, in a significant respect, also been underwritten by a widespread appeal to the general values of safety and security as overriding motivations and justifications for their use. While these values are undoubtedly crucial to consider in weighing the justifiability of any high-stakes technology, some critics have pushed back on such blanket invocations of safety and security, in the case of facial analysis technologies, as anxiety-mongering reflexes which tend to obscure the actual dynamics of subordination and control that influence the majority's impetus to adoption. They ask instead, "safety and security *for whom*?", and suggest that "the claim that biometric surveillance 'makes communities safer' is heavily marketed but loosely backed."[55]

Indeed, the prevailing opacity of the complex facial analysis technology supply chain makes answering the question of effectiveness, in most cases, all but impossible. From the oft-shadowy creation of largescale datasets[56] and the proprietary secrecy of these systems to the veiled character of surveillance practices, a degree of opaqueness would seem endemic in current contexts of innovation and use. As true as this may be, the obscurity of the development and deployment of facial analysis technologies is also not unrelated to the prioritisation of the appeal to safety and security itself. Public debates around the adoption of FDRTs are often framed through the generic lens of a necessary trade-off between values of safety and security, on the one side, and those of privacy and other individual and civil liberties, on the other (Figure 8).

However, the rhetoric of unavoidable trade-offs often combines with the treatment of the importance of safety and security as a kind of trump card of realism lying in wait. This wholesale privileging of safety and security can generate misleading and evasive narratives that insidiously secure innovation opacity in the name of exigency while, at the same time, drawing needed attention away from actionable steps that are available to safeguard transparency, accountability, appropriate public and individual consent, fundamental rights and freedoms, privacy protections, and wider impact awareness. Moreover, continual invocations of this sort of trade-off rhetoric can allow for the baseline assumption of the legitimacy of the use of facial analysis technologies to become entrenched thereby precluding informed and critical public dialogue on their very justifiability and ethical permissibility. This has already led to a focus on the *use-context* of these technologies in existing ethical guidance — with run-time and in-operation principles such as effectiveness, proportionality, necessity, and cost-effectiveness moved to centre stage – rather than a focus on the broader *ex ante* ethical concerns surrounding *use-justifiability* – with questions about the short- and long-term effects of these systems on individual self-development, autonomy, democratic agency, social cohesion, interpersonal intimacy and community wellbeing depreciated, set aside or shelved altogether.

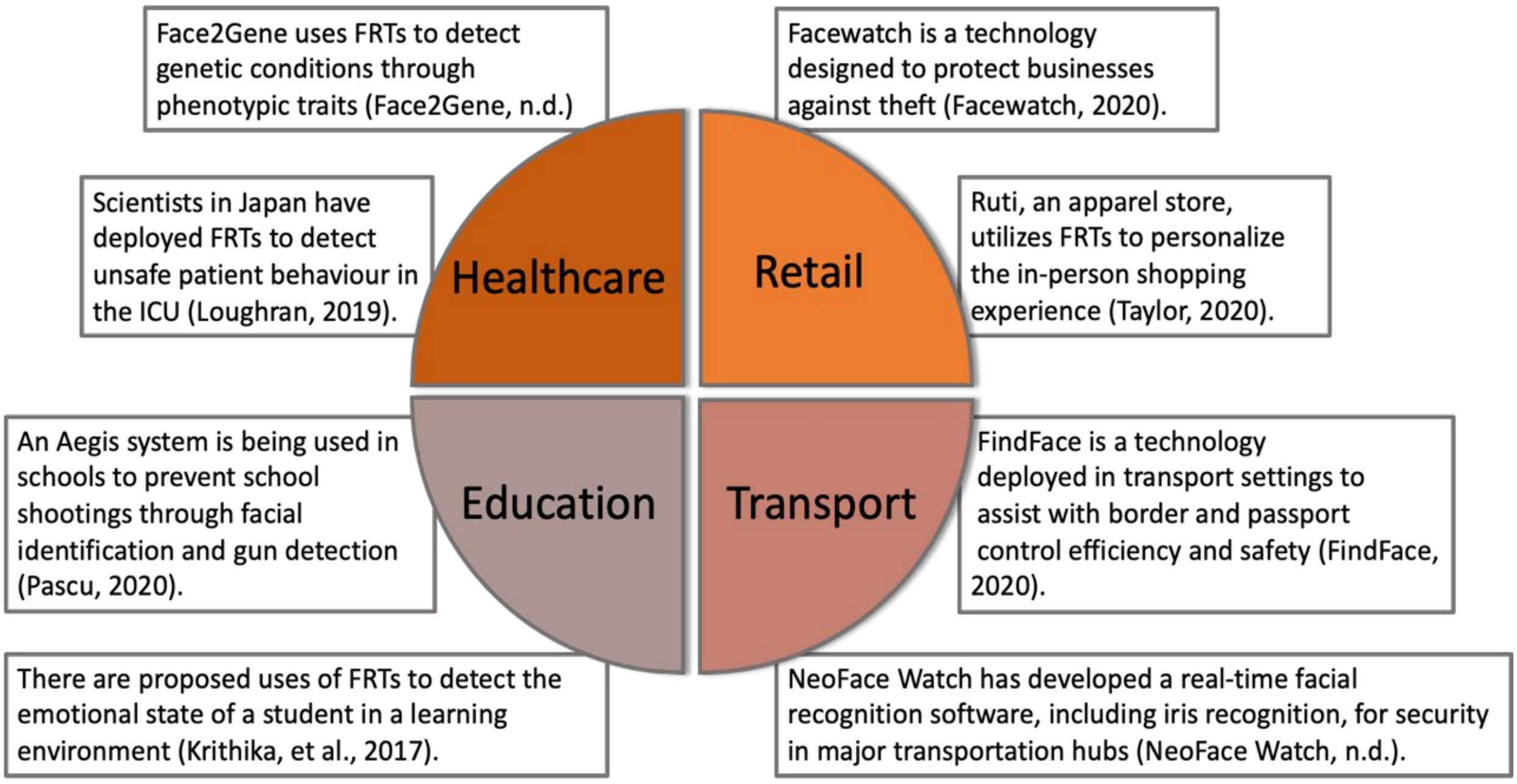

*Figure 8: Examples of Facial Recognition Technologies Employed Across Various Sectors*

Meanwhile the unregulated mass-peddling of facial surveillance systems by multiplying international vendors proceeds apace, fuelled by a corresponding rush to procure these tools by government agencies and companies that fear they will fall behind in the sprint to a non-existent technological finish line of total computational control, risk management and prediction. This has led many journalists, advocates and academics to send up warning flares about "the ever-creeping sprawl of face-scanning infrastructure."[57] Beyond intolerable discriminatory outcomes, this proliferation, it is feared, poses significant threats to the basic normative qualities and preconditions of a free and flourishing modern democratic society. A ubiquity of connected cameras and mobile recording devices that are linked to networked face databases will enable intrusive monitoring and tracking capacities at scale. These will likely infringe upon valued dimensions of anonymity, self-expression, freedom of movement and identity discretion as well as curtail legitimate political protest, collective dissent and the exercise of solidarity-safeguarding rights to free assembly and association.[58] Adumbrations of this are, in fact, already operative in the real-time surveillance of public events and protests in India, Hong Kong, the US and the UK creating tangible consternation about the chilling effects on democratic participation and social integration that hyper-personalised panoptical observation may have.[59]

Similar deleterious impacts are becoming progressively more evident in non-governmental applications of facial analysis technologies, where a growing number of private security outfits working in the retail, education, banking, housing, health and entertainment sectors are creating surveillance databases and monitoring infrastructures that may violate basic rights to consent and privacy, transgress civil liberties, human rights and due process protections and normalise pervasive behavioural scrutiny. For instance, in Portland, Oregon, a chain of petrol station convenience stores called Jacksons has, for the past two years, employed a facial recognition system that scans the faces of all night-time patrons to determine whether or not to grant them entry access to the shop. If a customer's face does not match any on their watchlist database of prohibited shoplifters and troublemakers, the doors are automatically unlocked and opened; otherwise entry is not permitted. This process is entirely non-consensual, non-transparent and requires the unchecked collection and processing of sensitive personal data (which are sent out-of-state to the company's private server in Idaho for analysis).[60] Though Jacksons will now have to wind down its use of this system in accordance with the Portland city council's new facial recognition ban, Blue Line Technology, the vendor of the application, is actively marketing the same product (with additional options to share watchlist databases among businesses) to housing complexes, hospitals, banks, and schools across the US.[61]

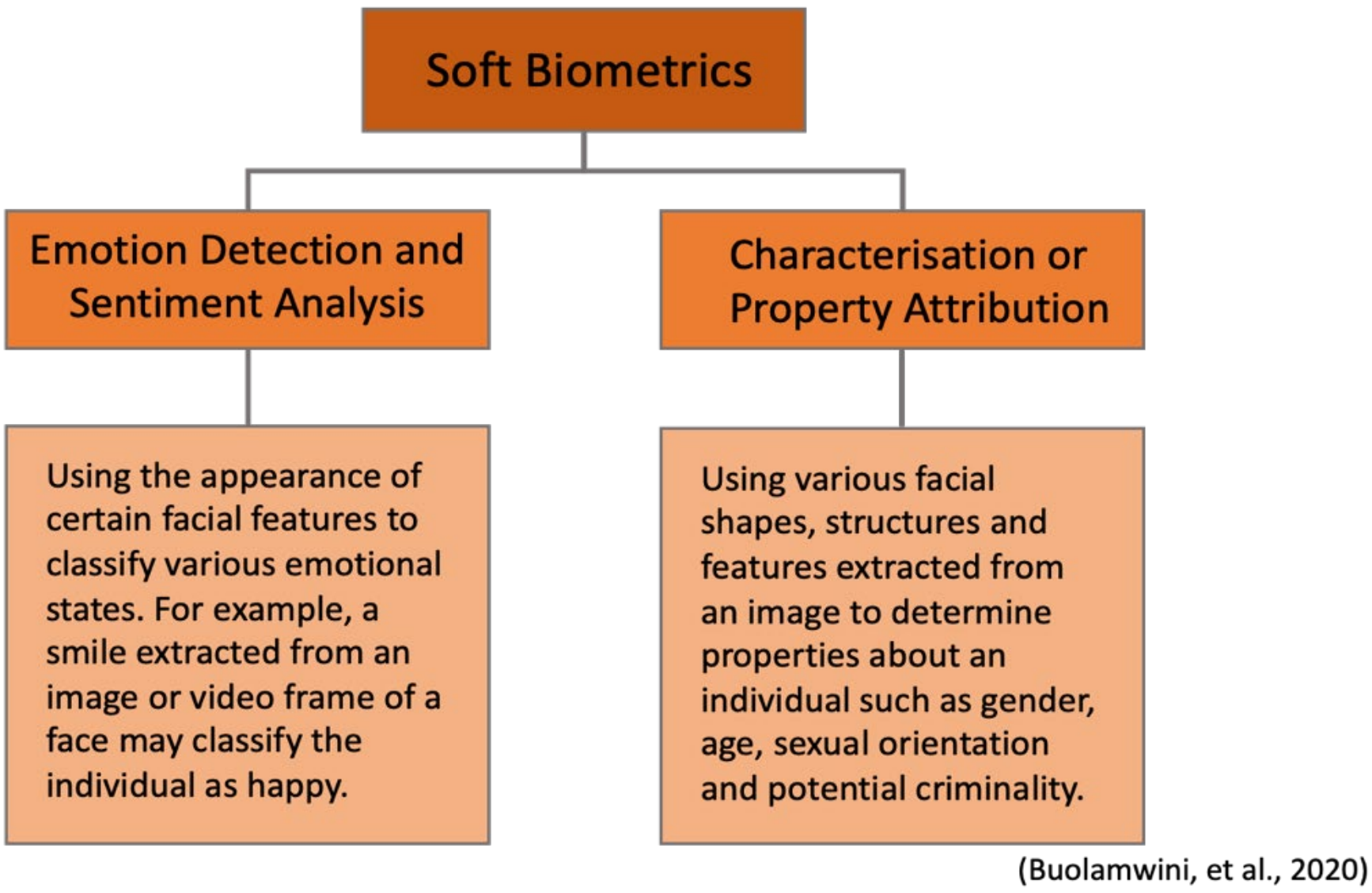

*Figure 9: Two Primary Uses of Soft Biometrics: Emotion Detection and Characterisation or Property Attribution*

As troubling as these indicators of the skulking colonisation of everyday rights and freedoms by face surveillance infrastructures may seem, they do not yet scratch the surface of the potential penetration into developmentally critical aspects of individual self-formation, socialisation and interpersonal connection that face-based technologies such as emotion detection and affect recognition may have (Figure 9). The exploding industry of computer vision systems that purport to detect emotional states from facial expressions now spans from areas like candidate vetting in job recruitment, personalised advertising and insurance fraud detection to workplace monitoring, attention-level/engagement assessment in classroom environments and deception analysis of criminal suspects in law enforcement.[62] Amazon's Rekognition facial analysis platform claims to effectively recognise eight emotions in tracked faces (happy, sad, angry, surprised, disgusted, calm, confused and fear), while IBM has patented an affect monitoring system for video conferences that provides "real-time feedback of meeting effectiveness" to presenters and moderators by reporting surveilled changes in the aggregate and individual sentiments of attendees.[63] More alarmingly, the US Department of Homeland Security is working on face analysis systems that infer the intent to harm from visual and physiological signals.[64]

Despite the wide-reaching but often concealed commercial employment of these tools of behavioural supervision and steering, critics have emphasised the dubious foundations of the entire endeavour of automated emotion detection and analysis. Drawing on scholarship in psychology, anthropology, and other social sciences, they have argued that much of face-based affect recognition is predicated on the false equation of the appearance of affective states in facial expressions (that have been generically or stereotypically classified and labelled by humans as representing given emotions) and the actual inner emotions that are experienced.[65] Not only are there cultural and intragroup variations in ways of emotive comportment—that is, in how one expresses fear, happiness, anger, etc. when these emotions are present within the self—but expressions of emotion are themselves also intimately connected to and influenced by the particular contexts, relationships and situations within which they arise. A smile or a scowl expressed in a social gathering or during a conversation, for example, may conform to culturally or relationally determined behavioural expectations or interactional norms rather than reflect an interior emotional state.

The shortcomings of data-driven emotion detection and affect recognition algorithms in managing these cultural and behavioural variations and contextual differences is directly linked to the set of ethical issues raised by their unbounded use. The epistemic and emotive qualities of interactive communication that enable human beings to pick up on and respond to subtle variations in the behavioural cues of others involve capacities for interpretation, empathic connection, common sense, situationally-aware judgment and contextual understanding. The facial recognition and emotion detection software that is used to pinpoint affect in facial expressions has none of these abilities. The mechanisms of statistical generalisation based upon which computational systems generate predictions about emotional states are categorically distinct from the intersubjective competences that allow humans to be able to interpret and understand each other's expressions, feelings and gestures. This difference in kind creates an ethically significant gap between the depersonalising character of computationally generated statistical inference—where quantified estimates of numerical patterns in data distributions are outputted under the guise of "recognising" emotion in a face—and the concrete reality of shared human experience – where the development of autonomous and properly socialised selves turns on the endurance of fluid interpretive relationships between people and on the moral demands placed on each person to respond to both cultural variation and to the differences that are native to the uniqueness and particularity of individual personalities. It is in relation to these imperatives of recognising the uniqueness of individual self-expression and of responding to the moral demands of difference that affect detection software unequivocally founders at both ethical and functional levels.

What is cultivated, instead, in a social world wherein automated face-based affect detection produces pervasive infrastructures of data-driven behavioural nudging, manipulation and control, is a crippling of the interpersonal dynamics of human self-creation. It is a reality in which the at-scale predictability of datafied subjects, who are algorithmically tempered to adhere behaviourally to the patterns of the past that have been trained into emotion recognition models, comes to progressively supplant the open and indeterminate character of agential experience that has defined and underwritten modern notions of freedom, flourishing and self-determination from the start. It is a reality, moreover, where the reciprocal character of action coordination through communicative participation and interactively-based value formation—the very bedrock of democratic forms of life—are gradually displaced by de-socialising mechanisms of automated behavioural regulation which operate on a totalising logic of normalisation and of the anticipatory pre-emption of uncertainty.

Such an impoverishment of individual and democratic agency and participatory self-creation evokes what many social thinkers downstream from Michel Foucault have called the disciplinary or carceral society (Figure 10). On the latter view, a world of pervasive surveillance and penetrative normalisation renders human subjects docile through regimes of constant

| Normalisation | Apparatus | Power/Knowledge | Disciplinary Society |
|---|---|---|---|
| Process by which concepts, standards and practices in society become the status quo and are seen as 'normal' and as parts of everyday life. This institutionalisation of norms establishes and structures shared meaning. | An ensemble of institutions, administrative techniques, regulatory decisions, scientific statements, knowledge structures and expert discourses that work in tandem to maintain and exercise power in society. | Foucault believed that power and knowledge are intertwined rather than disparate concepts. Power manifests via the enactment of accepted forms of knowledge, and through the shaping of knowledge, power is constituted. | A form of social life in which power organises human practices through the mechanisms of discipline, supervision and behavioural regulation found in institutions such as prisons. The presence of constant surveillance leads to omnipresent disciplining and subjection. |

(Foucault, 1976)

*Figure 10: Key Foucault Concepts*

observation, measurement, supervision, correctional practice and self-discipline.[66]

As Foucault would say, disciplinary power identifies deviants who skew from the prevailing and authoritative field of normalcy, so that it can control and neutralise them. The spaces of knowledge, apparatuses, procedures and techniques through which this type of power is enacted in the social world form and norm passive subjectivities instead.

As one such subset of disciplinary technologies, emotion detection algorithms—and their sibling face-based trait-attribution models that questionably claim to infer properties like criminality and sexual orientation from facial morphology[67]—can be seen, along Foucauldian lines, to operate as instruments of power/knowledge and subjugation. In them, physical mechanisms of expression-capture or physiognomic measurement are taken as techniques of knowledge production (within a wider field statistical expertise) that yield a scientific grasp on the inner states or properties of observed individuals. The "normalising gaze" of this kind of computer vision tool, to use Foucault's words, "manifests in the subjection of those who are perceived as objects and the objectification of those who are subjected."[68] All in all, in a disciplinary society where tools of normalisation like these become omnipresent, individual subjects are subjugated as objects of authoritative knowledge and thereby become sitting targets of carceral regulation and scientific management. Distressingly, however, the potential ubiquity of face scanning infrastructures would seem, in an important respect, to supersede even this dystopic Foucauldian vision of the disciplinary society. For, Foucault held out the hope that, in the carceral society, liberation and emancipatory transformation would remain possible inasmuch as there would still exist the potential for dynamics of deviance and critical forces of resistance to crack through the shell of normalisation and to bring about transformative and liberating outcomes that broaden human freedom. Because, on Foucault's understanding, power is depersonalised and structurally dispersed, it leaves fissures and breaches in the disciplinary order that enable radical transgressions into the novelty of the indeterminate present of human self-creation. Indeed, the very logic of disciplinary society necessarily presupposes dimensions of deviation, resistance and emancipation devoid of which there would be no need for discipline to begin with.

To the contrary, in a society driven toward the ubiquitous predictability of datafied subjects—a prediction society—hyper-personalising anticipatory calculation pre-empts deviation and resistance as such and at its embodied source. The impetus to annihilate otherness apiece (whether in the form of the uniqueness of individual identity or sociocultural difference) is, in fact, the defining feature. Here, the proliferation of panoptical systems that predict and manipulate individual traits and inner states of intention and emotion stem the very possibility of anomaly and deviation in virtue of the eradication, or at least the effective domestication, of uncertainty *per capita*. Through the tailored algorithmic management of the inner life and individual characteristics of each person, predictory power singles out the behavioural mind and body in order to create consummately calculable subjectivities. In the prediction society, this exercise of hyper-targeted but wholly depersonalised pre-emptory power hinges on the denial of the precise dynamics of deviation and difference upon which the disciplinary society is still intrinsically parasitic.

Such a horror story about the possible emergence of a prediction society in which essential features of the humanity of the human go missing is meant to be cautionary tale. This sort of anticipatory view of how the unchecked development of pervasive face-based surveillance and emotion and trait detection infrastructures could lead to potential unfreedom should, however, also be seen as a critical, forward thinking exercise—a strategy that is meant to help us reflect soberly on what is ethically at stake in the current pursuit of this kind of high impact data-driven innovation. Prospects of a social world shorn of the qualities of democratic agency, individual flourishing and interpersonal connection that have afforded the realisation of some of the basic features of modern identity, and of its corresponding expansions of social freedom and of individual and civil liberties, should concern us greatly. Such prospects should be reason enough to interrupt the seamless advancement of these sorts of technologies so that open and inclusive dialogue among affected individuals and communities about their justifiability and utility can occur. After all, it is ultimately those whose interests and lives are impacted by potentially deleterious paths of innovation rather than those who benefit financially or politically from marching down them that should set the direction of travel for societally, globally and, indeed, intergenerationally ramifying technological transformation.

## Conclusion

At bottom, the demand for members of society writ large to come together to evaluate and reassess the ethical permissibility of FDRTs originates in the acknowledgment that their long-term risks to the sustainability of humanity as we know it may so far outweigh their benefits as to enjoin unprecedented technological self-restraint. Some scholars, emphasising the toxic racialising and discriminatory effects of these systems, have likened them to plutonium. They have pointed out that, like nuclear waste, facial recognition algorithms are “something to be recognised as anathema to the health of human society, and heavily restricted as a result.”[69] Others have stressed how the harmful effects of these systems may manifest in the universal loss of personal privacy, relational intimacy and capacities for spontaneity and uncoerced identity formation.
They have described these technologies as "a menace disguised as a gift" and called for an outright prohibition on them so that rudimentary possibilities for human flourishing can be safeguarded.[70]

Above all else, these stark warnings open up to public dialogue and debate a crucial set of ethical questions that should place the fate of facial recognition technologies in the lap of the moral agency of the present. The democratically-shaped articulation of the public interest regarding the further development and proliferation of these systems should, in no small measure, steer the gathering energies of their exploration and advancement. It is only through inclusive deliberation and the community-led prioritisation of the values and beliefs behind innovation trajectories that societal stakeholders will come to be able to navigate the course of technological change in accordance with their own shared vision of a better, more sustainable and more humane future.

Still, confronted with the immediacy of the noxious practices already present, contemporary society faces a second, remedial task—the task of taking the wheel of technology governance from those heedless, opportunistic and ethically-blindfolded technologists who are building FDRTs without due attention to responsible innovation practices and wider impacts. At minimum, this will involve mandating the following aspects of ethical technology production and use:

**Robust governance mechanisms in place to secure transparency and accountability across the entire design, development and deployment workflow**: Practices of building and using facial recognition technologies must be made accountable and transparent *by design.* A *continuous chain of human responsibility* must be established and codified across the whole production and use lifecycle, from data capture, horizon-scanning and conceptualisation through development and implementation. This applies equally to the vendors and procurers of such technologies, who must cooperate to satisfy such requisites of answerable design and use. It also applies to the collection and curation practices that are involved in building and using largescale face datasets. All of these processes must be made appropriately traceable and auditable from start to finish. Facial recognition technology builders and users must be able to demonstrate that their design practices are responsible, safe, ethical and fair from beginning to end and that their model's outputs are justifiable. This involves:

(1) Maintaining strong regimes of professional and institutional transparency by availing to the public clear and meaningful information about the facial recognition technologies that are being used, how these have been developed and how they are being implemented;

(2) Putting in place deliberate, transparent and accessible process-based governance frameworks so that explicit demonstration of responsible innovation practices can help to assure the safety, security, reliability and accuracy of any given system as well as help to justify its ethical permissibility, fairness and trustworthiness;[71]

(3) Establishing well-defined auditability trails through robust activity logging protocols that are consolidated digitally in process logs or in assurance case documentation and clearly conveyed through public reporting, during third-party audits and, where appropriate, in research publications;

(4) Clarifying specific outcomes of the use of any given system to impacted individuals or their advocates by clearly and understandably explaining the reasoning behind particular results in plain and non-technical language.[72]

**Robust guarantees of end-to-end privacy preservation, individual and community consent, and meaningful notice in accordance with fundamental rights and freedoms, the prohibition of discrimination and data protection principles:** Practices of building and using facial recognition technologies must incorporate safeguards that ensure privacy preservation, unambiguous and meaningful notice to affected parties and sufficient degrees of individual and community consent. This applies equally to (1) the data capture, extraction, linking and sharing practices behind the production of largescale face datasets and benchmarks as well as the utilisation of these or any other data in model training, development, auditing, updating and validation and (2) the "in the wild" system deployment practices and runtime data collection and processing activities in both public and private spaces. Across both of these dimensions of the facial recognition technology supply chain, due attention must be paid to:

(1) Securing individual rights to privacy that accord with the reasonable expectations of impacted individuals given the contexts of collection, processing and use—endeavours to define such reasonable expectations must include considerations of rights to anonymity and identity discretion, freedom from the transformative effects of intrusive data collection and processing on the development of individual personality, and rights to the preservation of the confidentiality of chosen relationships;

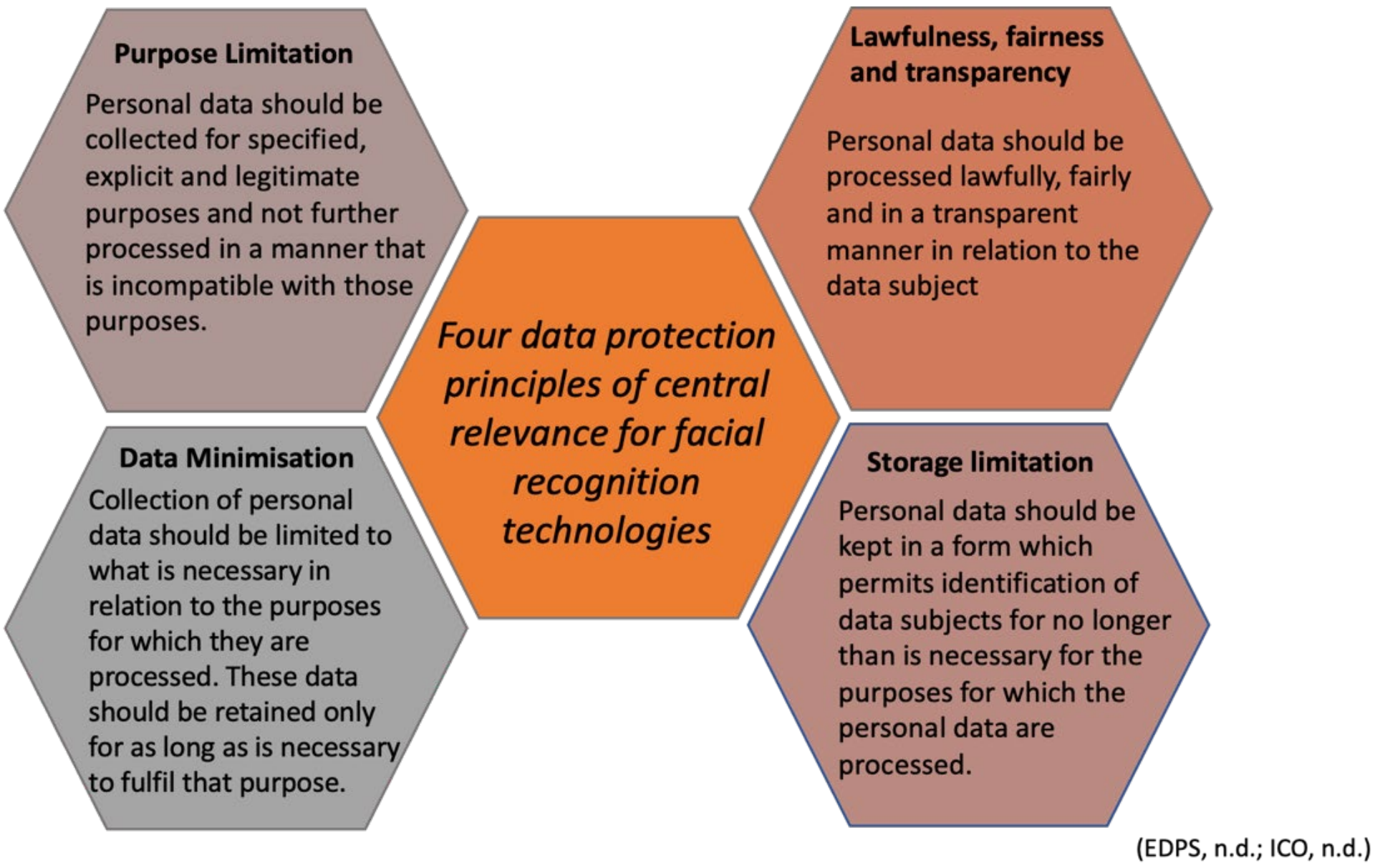

*Figure 11: Four Primary Data Protection Principles*

(2) Securing the affirmative consent of affected individuals (where legal exceptions are not present) in advance of any data collection, processing and retention that is related to the development and deployment of facial recognition technologies and the compilation of facial databases;[73]

(3) Balancing the onus of individual consent and information control[74] with diligent organisation-level assessment of and conformity with applicable human rights and data protection principles (including purpose limitation, proportionality, necessity, non-discrimination, storage limitation, and data minimisation) (Figures 11 and 12);[75]

(4) Building public consent by involving affected communities in the evaluation of the purpose, necessity, proportionality and effectiveness of specific applications as well as in the establishment of their consonance with the public interest;

(5) Tailoring notice about the deployment of a given system in an application-specific way that meaningfully informs affected individuals of its intended use as well as of its purpose and how it works;[76]

(6) Designing notice and disclosure protocols that enable and support due process and actionable recourse for impacted individuals in the event that harmful outcomes occur.[77]

**Robust measures to secure comprehensive bias-mitigation measures and discrimination-aware design, benchmarking and use:** Practices of building and using facial recognition technologies must incorporate discrimination-aware strategies for bias-mitigation comprehensively and holistically—addressing both the technical challenges of mitigating the biases that originate in unbalanced samples or that are baked into datasets and the sociotechnical challenges of redressing biases that lurk

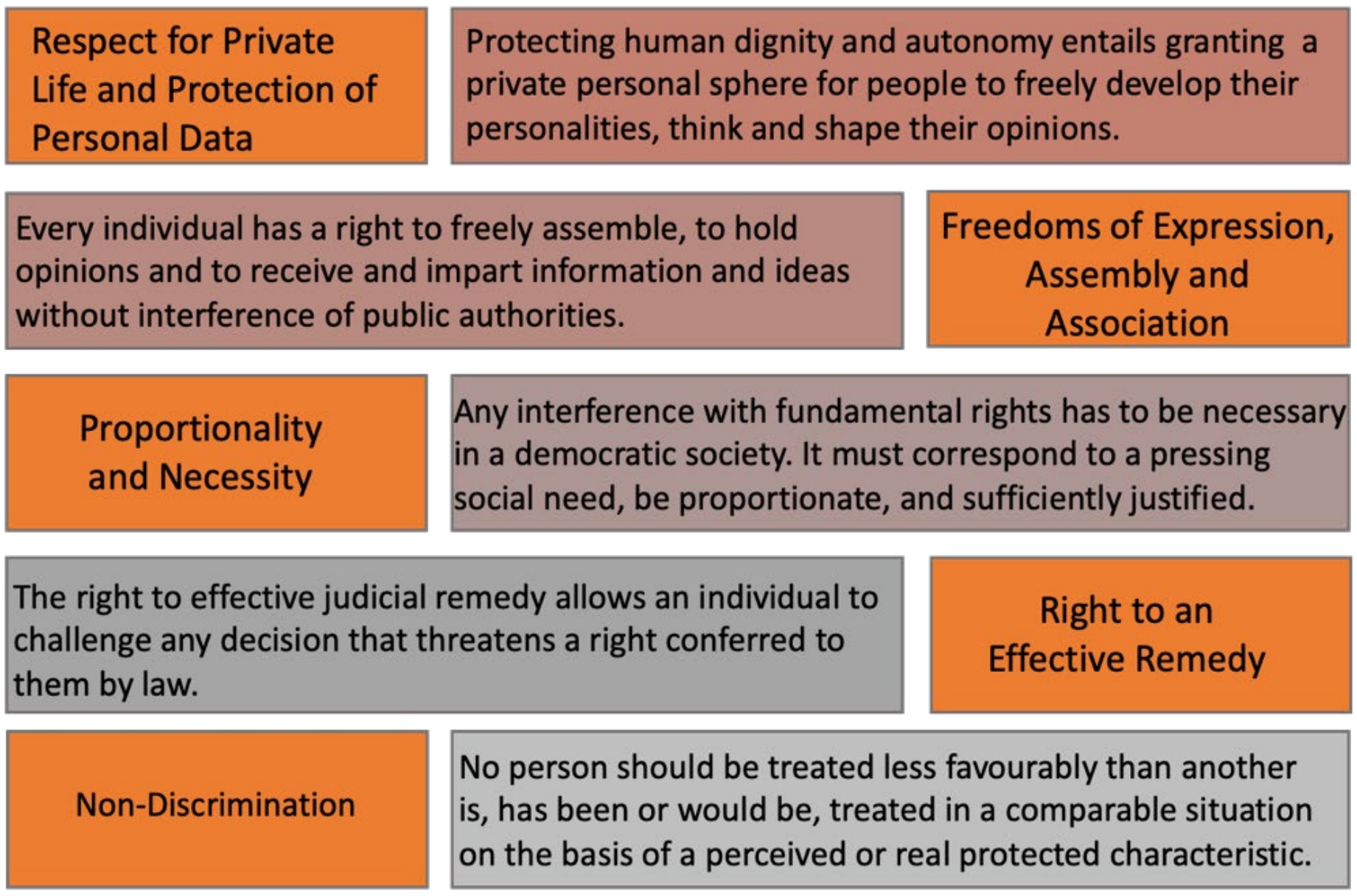

(FRA, 2020)

*Figure 12: Human Rights Data Provisions Related to FRTs*

within the design, development and deployment practices themselves. Examples of bias mitigation techniques can be found in Table 2.

Attention must be focused on:

(1) Understanding and rectifying representational imbalances in training datasets, so that their influence on differential performance can be mitigated;

(2) Engaging in proactive bias-mitigation to address other technical issues in computer vision processing (such as confounding environmental covariates like illumination) that lead to differential performance beyond sampling biases and imbalanced datasets;

(3) Engaging in inclusive, deliberative and bias-aware labelling and annotation practices that make explicit how sensitive demographic and phenotypic features have been classified, aggregated, separated and categorised;

(4) Engaging, at the very inception of facial recognition project development, in inclusive and community-involving processes of horizon scanning, agenda-setting, problem formulation and impact assessment, so that those groups most affected by the potential use of the system can participate in the setting the direction for the innovation choices that impact them;

(5) Constructing and utilising benchmarks that support the measurement of differential performance and that enable clear and explicit reporting of system limitations and accuracy differences as these are distributed among relevant phenotypic and demographic subgroups, while also making explicit potential performance-influencing gaps between benchmarking processes and real-world conditions of application deployment;

(6) Training users and implementers of facial recognition technologies in a bias-aware manner that imparts an - understanding of the limitations of these systems (and statistical reasoning, more generally) as well as of the cognitive and discriminatory biases that may lead to over-reliance and over-compliance, on the one hand, and result-dismissal or neglect, on the other;

(7) Setting in place explicit monitoring protocols that foster the regular assessment and re-assessment of potential discriminatory harms or impacts which may occur once a facial recognition system is live and in operation.

*Table 2: Examples of Bias Mitigation Techniques in Facial Detection and Recognition Research*

| Title | Author(s) | Year | Motivation | Method |
|---|---|---|---|---|
| Human age estimation: What is the influence across race and gender? | Guodong Guo and Guowang Mu | 2010 | Authors determined that when “crossing” gender and race there is an increasing amount of errors related to age classification. | Dynamic classification was used on selected demographics to improve model performance. |
| Face Recognition Performance: Role of Demographic Information | Brendan Klare, Mark Burge, Joshua Klontz, Richard Vorder Bruegge, and Anil Jain | 2012 | There are varying accuracies across cohorts when isolated based on age, race/ethnicity, and gender. | The authors observed improved accuracy metrics for age and race/ethnicity cohorts by training model on a specific cohort. This is termed dynamic face matcher selection. Various facial recognition models each trained on a specific cohort are available for the system to draw on, improving face recognition accuracy. |
| One-shot Face Recognition by Promoting Underrepresented Classes | Yandong Guo and Lei Zhang | 2018 | Many training datasets contain class imbalances. | The authors created a new feature extraction model with a corresponding multi-class classifier using “underrepresented-classes promotion loss.” This function standardizes weight vectors across all classes to ensure underrepresented groups are present in the model training process. |
| InclusiveFaceNet: Improving Face Attribute Detection with Race and Gender Diversity | Hee Jung Ryu, Hartwig Adam, and Margaret Mitchell | 2017 | Various face attribute detection models have significant model performance differences across race and gender. | InclusiveFaceNet was created to detect face attributes through the use of a public dataset that contains learned race and gender identities. Additionally, the model withholds demographic inference during the face attribute detection component of the model, thereby maintaining demographic privacy. |
| The Impact of Age and Threshold Variation on Facial Recognition Algorithm Performance Using Images of Children | Dana Michalski, Sau Yee Yiu, and Chris Malec | 2018 | There is a small area of literature surrounding algorithmic performance in models containing images of children with respect to age and age variation. | A comparison of threshold variation and fixed threshold approaches resulted in significant differences in model performance. The authors concluded that threshold variation may be useful in facial recognition involving images of children. |
| Turning a Blind Eye: Explicit Removal of Biases and Variation from Deep Neural Network Embeddings | Mohsan Alvi, Andrew Zisserman, and Christoffer Nellaker | 2018 | Neural networks can encode biases that are present within the training data, especially when a training dataset is imbalanced. | The authors created an algorithm that removes biases from feature embeddings. This algorithm is proven to be especially useful when dealing with extremely biased datasets. |
| Uncovering and Mitigating Algorithmic Bias through Learned Latent Structure | Alexander Amini, Ava Soleimany, Wilko Schwarting, Sangeeta Bhatia, and Daniela Rus | 2019 | Facial recognition algorithms have demonstrated biased results, especially towards underrepresented groups. | The authors have designed an algorithm for racial and gender bias mitigation through the use of an autoencoder that utilizes learned latent distributions to calculate re-weightings for data points in the dataset. |
| An Experimental Evaluation of Covariates Effects on Unconstrained Face Verification | Boyu Lu, Jun-Cheng Chen, Carlos Castillo, and Rama Chellappa | 2018 | Covariates can negatively impact the model performance of facial verification algorithms. | Noise was removed from the dataset resulting in a curated dataset that increased overall model performance and decreased false acceptance rates. |

Table expanded from (Drozdowski, et al., 2020)

Notwithstanding these concrete recommendations, those engaging in the sufficiently discrimination-aware design and deployment of FDRTs must also continually position their development- and use-decisions in the wider context of historical and structural injustices and systemic racism. The breakneck speed of the current growth in the use of live facial recognition and face matching databases for identification raises issues about the haphazard compounding of existing social inequities and dynamics of oppression.

Harms arising from the deeply entrenched historical patterns of discrimination that are ingrained in the design and use of FDRTs now threaten to converge with the broader injurious impacts that societally consequential algorithmic systems like predictive risk modelling in law enforcement and social services have already had on marginalised populations who have long been subjected to over-surveillance and over-policing. The potential for digital epidermalisation and pseudo-scientific racialisation to further entrench intolerable social injustices must remain a primary consideration in every potential project to use facial recognition technologies.

[1] As a convention, I will use this abbreviation (FDRTs) interchangeably with the phrases "facial detection and analysis" models, algorithms, technologies and systems. All of these terms are meant to encompass the full range of facial processing technologies.

[2] Ming-Hsuan Yang, David J. Kriegman, and Narendra Ahuja, "Detecting Faces in Images: A Survey," *IEEE Transactions on Pattern Analysis and Machine Intelligence* 24, no.1 (2002): 34-58.

[3] Paul Viola and Michael Jones, "Rapid Object Detection Using a Boosted Cascade of Simple Features," *Proceedings of the 2001 IEEE Computer Society Conference on Computer Vision and Pattern Recognition*, (2001), https://doi.org/10.1109/cvpr.2001.990517

[4] Kate Crawford and Trevor Paglen, "Excavating AI: The Politics of Training Sets for Machine Learning," (September 19, 2019), https://excavating.ai

[5] Richard Dyer, *White: Essays on Race and Culture* (London, UK: Routledge, 1996).

[6] Ruha Benjamin, *Race after Technology: Abolitionist Tools for the New Jim Code* (Cambridge, UK: Polity Press, 2019).; Simone Browne, *Dark Matters: on the Surveillance of Blackness* (Durham, NC: Duke University Press, 2015).; Dyer, *White*.; Sarah Lewis, "The Racial Bias Built into Photography," *The New York Times*, April 25, 2019.; Lorna Roth, "Looking at Shirley, the Ultimate Norm: Colour Balance, Image Technologies, and Cognitive Equity," *Canadian Journal of Communication* 34, no. 1 (2009), https://doi.org/10.22230 /cjc.2009v34n1a2196 ; Lorna Roth, "Making Skin Visible through Liberatory Design," in *Captivating Technology: Race, Carceral Technoscience, and Liberatory Imagination in Everyday Life,* ed. Ruha Benjamin (Durham, NC: Duke University Press, 2019).

[7] Roth, "Looking at Shirley," 118.

[8] Lewis, "The Racial Bias Built into Photography," April 25, 2019.

[9] Roth, "Looking at Shirley."

[10] Syreeta McFadden, "Teaching the Camera to See My Skin," *BuzzfeedNews*, April 2, 2014, https://www.buzzfeednews.com/article/syreetamcfadden/teaching-the-camera-to-see-my-skin

[11] Browne, *Dark Matters.*

[12] Ruha, *Race after Technology*.

[13] Gwen Sharp, "Nikon Camera Says Asians are always Blinking," *TheSocietyPages.org*, May 29, 2009, https://thesocietypages.org/socimages/2009/05/29/nikon-camera-says-asians-are-always-blinking/

[14] David Hankerson, et al., "Does technology have race?" *Proceedings of the 2016 CHI Conference Extended Abstracts on Human Factors in Computing Systems,* (2016): 473-486.

[15] Brian X. Chen, "HP Investigates Claims of 'Racist' Computers," *Wired.com,* December 22, 2009, https://www.wired.com/2009/12/hp-notebooks-racist/

[16] Aarti Shah, "HP Responds to 'Racist' Video on Blog," *PRweek.com*, December 22, 2009, https://www.prweek.com/article/1269196/hp-responds-racist-video-blog

[17] Aarti Shah, "HP Responds to 'Racist' Video on Blog."

[18] For further discussion of sociotechnical issues surrounding racism in software design, see Christian Sandvig, Kevin Hamilton, Karrie Karahalios, and Cedric Langbort, "When the Algorithm itself is a Racist: Diagnosing Ethical Harm in the Basic Components of Software," *International Journal of Communication* 10(2016), 4872-4990.

[19] P. Jonathan Phillips, Patrick Grother, Ross Micheals, Duane Blackburn, Elham Tabassi, and Mike Bone, "Face recognition vendor test 2002: evaluation report," National Institute of Standards and Technology, NISTIR, vol. 6965, 2003; Patrick Grother, G. W. Quinn, and P. Jonathan Phillips, "Report on the evaluation of 2D still-image face recognition algorithms," National Institute of Standards and Technology, Tech. Rep.

NISTIR 7709, August 2010; Mei Ngan, Mei Ngan, and Patrick Grother. Face recognition vendor test (FRVT) performance of automated gender classification algorithms," National Institute of Standards and Technology, 2015. Patrick Grother, Mei Ngan, and K. Hanaoka, "Ongoing face recognition vendor test (FRVT) part 3: Demographic effects," National Institute of Standards and Technology, Tech. Rep. NISTIR 8280, December 2019.

[20] P Jonathon Phillips, Fang Jiang, Abhijit Narvekar, Julianne Ayyad, and Alice J O'Toole, "An other-race effect for face recognition algorithms," *ACM Transactions on Applied Perception* (*TAP*), 8(2) (2011): 14.; Brendan F Klare, Mark J Burge, Joshua C Klontz, Richard W Vorder Bruegge, and Anil K Jain, "Face recognition performance: Role of demographic information," *IEEE Transactions on Information Forensics and Security*, 7(6) (2012): 1789–1801.

[21] Joy Buolamwini and Timnit Gebru, "Gender shades: Intersectional accuracy disparities in commercial gender classification," *Conference on fairness, accountability and transparency* (2018): 77-91.

[22] Michele Merler Nalini Ratha, Rogerio S. Feris, and John R. Smith, "Diversity in faces," *arXiv preprint* (2019), arXiv:1901.10436.

[23] Michele Merler, et al., "Diversity in faces."

[24] Benjamin Wilson, Judy Hoffman, and Jamie Morgenstern. "Predictive inequity in object detection," *arXiv preprint* (2019) arXiv:1902.11097.

[25] Crawford and Paglen, "Excavating AI: The Politics of Training Sets for Machine Learning."

[26] Luke Stark, "Facial Recognition Is the Plutonium of AI," *XRDS: Crossroads, The ACM Magazine for Students* 25, no. 3 (October 2019): pp. 50-55, https://doi.org/10.1145/3313129.

[27] Browne, *Dark Matters.*

[28] For further discussion see: Morgan Klaus Scheuerman, Jacob M. Paul, and Jed R. Brubaker, "How Computers See Gender: An Evaluation of Gender Classification in Commercial Facial Analysis Services," *Proceedings of the ACM on Human-Computer Interaction 3,* no. CSCW (November 2019): 1–33, https://doi.org/10.1145/3359246.

[29] Kashmir Hill, "The Secretive Company that Might End Privacy as We Know It," *The New York Times*, January 18, 2020, https://www.nytimes.com/2020/01/18/technology/clearview-privacy-facial-recognition.html; Rebecca Heilweil, "The World's Scariest Facial Recognition Company, Explained," *Vox.com*, May 8, 2020, https://www.vox.com/recode/2020/2/11/21131991/clearview-ai-facial-recognition-database-law-enforcement; Ben Gilbert, "Facial-Recognition Software Fails to Correctly Identify People '96% of the Time,' Detroit Police Chief Says," *BusinessInsider.com,* June 30, 2020, https://www.businessinsider.com/facial-recognition-fails-96-of-the-time-detroit-police-chief-2020-6?r=US&IR=T

[30] Clare Garvie, Alvaro M. Bedoya, and Jonathan Frankle, "The Perpetual Line-Up: Unregulated Face Recognition in America," *Georgetown Law Center on Privacy & Technology*, (October 18, 2016): 25, 35, https://www.perpetuallineup.org/report

[31] Jameson Spivack and Clare Garvie, "A Taxonomy of Legislative Approaches to Face Recognition in the United States" in *Regulating Biometrics: Global Approaches and Urgent Questions*, ed. Amba Kak (AI Now Institute, 2020).

[32] Courtney Linder, "Why Portland Just Passed the Strictest Facial Recognition Ban in the US," *PopularMechanics.com,* September 12, 2020, https://www.popularmechanics.com/technology/security/a33982818/portland-facial-recognition-ban/

[33] Charlie Warzel, "A Major Police Body Cam Company Just Banned Facial Recognition," *The New York Times,* June 27, 2019, https://www.nytimes.com/2019/06/27/opinion/police-cam-facial-recognition.html

[34] Notably, though, both companies remain profitably interwoven in the police surveillance technology supply chain through their partnerships with smaller vendors and provision of critical cloud computing infrastructure. See: Michael Kwet, "The Microsoft Police State: Mass Surveillance, Facial Recognition, and the Azure Cloud," *TheIntercept.com,* July 14, 2020, https://theintercept.com/2020/07/14/microsoft-police-state-mass-surveillance-facial-recognition/

[35] Rebecca Heilweil, "Big Tech Companies Back Away from Selling Facial Recognition Technology to Police. That's Progress," *Vox.com,* June 11, 2020, https://www.vox.com/recode/2020/6/10/21287194/amazon-microsoft-ibm-facial-recognition-moratorium-police; Rebecca Heilweil, "Why it Matters that IBM is Getting Out of the Facial Recognition Business," *Vox.com,* June 10, 2020, https://www.vox.com/recode/2020/6/10/21285658/ibm-facial-recognition-technology-bias-business

[36] Bridges v. Chief Constable of South Wales police*, 2020 EWCA Civ 1058*, C1/2019/2670. https://www.judiciary.uk/wp-content/uploads/2020/08/R-Bridges-v-CC-South-Wales-ors-Judgment-1.pdf

[37] Nancy Fraser and Axel Honneth, *Redistribution or Recognition?: A Political-Philosophical Exchange*. (London, UK: Verso, 2003).

[38] "AI Now Report 2018" (The AI Now Institute, December 2018), https://ainowinstitute.org/AI_Now_2018_Report.pdf.

[39] "AI Now Report 2018."

[40] Adam Vaughan, "UK Launched Passport Photo Checker it Knew Would Fail with Dark Skin," *NewScientist.com,* October 9, 2019, https://www.newscientist.com/article/2219284-uk-launched-passport-photo-checker-it-knew-would-fail-with-dark-skin/

[41] @elainebabey, Twitter post, February 25, 2020, 10:56am, https://twitter.com/elainebabey/status/1232333491607625728

[42] Vaughan, "UK Launched Passport Photo Checker it Knew Would Fail with Dark Skin."

[43] Kashmir Hill, "Wrongfully Accused by an Algorithm," *The New York Times,* June 24, 2020, https://www.nytimes.com/2020/06/24/technology/facial-recognition-arrest.html

[44] Kashmir Hill, "Wrongfully Accused by an Algorithm."

[45] Kashmir Hill, "Wrongfully Accused by an Algorithm."

[46] Kashmir Hill, "Wrongfully Accused by an Algorithm."

[47] George Hunter, "Project Green Light to Add Facial Recognition Software," *The Detroit News,* October 30, 2017, https://eu.detroitnews.com/story/news/local/detroit-city/2017/10/30/detroit-police-facial-recognition-software/107166498/

[48] George Hunter, "Project Green Light to Add Facial Recognition Software."

[49] Elisha Anderson, "Controversial Detroit Facial Recognition Got Him Arrested for a Crime He Didn't Commit," *Detroit Free Press,* July 12, 2020, https://eu.freep.com/story/news/local/michigan/detroit/2020/07/10/facial-recognition-detroit-michael-oliver-robert-williams/5392166002/; Natalie O'Neill, "Faulty Facial Recognition Led to His Arrest-Now He's Suing," *Vice.com,* September 4, 2020, https://www.vice.com/en_us/article/bv8k8a/faulty-facial-recognition-led-to-his-arrestnow-hes-suing

[50] Oliver v. Bussa, Cassini, & City of Detroit, *State of Michigan*, September 4, 2020, https://www.documentcloud.org/documents/7202332-Draft-of-Michael-Oliver-Complaint-changes932020-1.html

[51] “Skin Colour in the Photo Checking Service,” *WhatDoTheyKnow*, July 30, 2019, https://www.whatdotheyknow.com/request/skin_colour_in_the_photo_checkin#incoming-1443718; Vaughan, “UK Launched Passport Photo Checker it Knew Would Fail with Dark Skin.”
[52] Vaughan, “UK Launched Passport Photo Checker it Knew Would Fail with Dark Skin.”
[53] “About FACER2VM,” *FACER2VM*, 2018, https://facer2vm.org/about/; Sebastian Klovig Skelton, “UK Facial Recognition Project to Identify Hidden Faces,” March 18, 2020, https://www.computerweekly.com/news/252480233/UK-facial-recognition-project-to-identify-hidden-faces
[54] Paul Wiles, “Biometrics Commissioner’s Response to the Home Office Biometrics Strategy,” *GOV.UK,* June 28, 2018, https://www.gov.uk/government/news/biometrics-commissioners-response-to-the-home-office-biometrics-strategy; Surveillance Camera Commissioner and Tony Porter, “Surveillance Camera Commissioner’s statement: Court of Appeal judgement (R) Bridges v South Wales Police – Automated Facial Recognition,” *GOV.UK,* August 11, 2020, https://www.gov.uk/government/speeches/surveillance-camera-commissioners-statement-court-of-appeal-judgment-r-bridges-v-south-wales-police-automated-facial-recognition
[55] Amba Kak, ed., “Regulating Biometrics: Global Approaches and Urgent Questions,” *AI Now Institute*, September 1, 2020, 11 https://ainowinstitute.org/regulatingbiometrics.html
[56] Vinay Uday Prabhu and Abeba Birhane, "Large image datasets: A pyrrhic win for computer vision?." *arXiv preprint,* (2020), arXiv:2006.16923.
[57] Evan Selinger and Woodrow Hartzog, “Letter to Assemblyman Chau,” April 13, 2020, https://www.aclunc.org/sites/default/files/2020.04.13%20-%20Scholars%20Letter%20Opp.%20AB%202261.pdf
[58]Browne, *Dark Matters*; Evan Selinger and Woodrow Hartzog, “Obscurity and Privacy,” *Routledge Companion to Philosophy of Technology* (Joseph Pitt and Ashley Shew, eds., 2014 Forthcoming), https://ssrn.com/abstract=2439866; Woodrow Harzog, “Privacy and the Dark Side of Control,” *iai news,* September 4, 2017, https://iai.tv/articles/privacy-the-dark-side-of-control-auid-882; Evan Selinger and Woodrow Hartzog, “What Happens When Employers Can Read Your Facial Expressions?” *The New York Times,* October 17, 2019, https://www.nytimes.com/2019/10/17/opinion/facial-recognition-ban.html; Maria Cristina Paganoni, “Ethical Concerns over Facial Recognition Technology,” *Anglistica AION an Interdisciplinary Journal* 23, no.1 (2019): 83-92.; Selinger and Hartzog, “Letter to Assemblyman Chau.”
[59] Kak, ed., “Regulating Biometrics: Global Approaches and Urgent Questions.”
[60] Everton Bailey, Jr., “Portland Considering Strictest Ban on Facial Recognition Technology in the U.S.,” *The Oregonian,* February 21, 2020, https://www.oregonlive.com/portland/2020/02/portland-considering-strictest-ban-on-facial-recognition-technology-in-the-us.html
[61] [61] Everton Bailey, Jr., “Portland Considering Strictest Ban on Facial Recognition Technology in the U.S..”
[62] Charlotte Gifford, “The Problem with Emotion-Detection Technology,” *TheNewEconomy.com,* June 15, 2020, https://www.theneweconomy.com/technology/the-problem-with-emotion-detection-technology
[63] Jayne Williamson-Lee, “Amazon's A.I. Emotion-Recognition Software Confuses Expressions for Feelings,” *OneZero.Medium.com,* February 12, 2020, https://onezero.medium.com/amazons-a-i-emotion-recognition-software-confuses-expressions-for-feelings-53e96007ca63 ; *Amazon Web Services,* “Amazon Rekognition Improves Face Analysis,” August 12, 2019, https://aws.amazon.com/about-aws/whats-new/2019/08/amazon-rekognition-improves-face-analysis/ ; Hernan Cunico and Asima Silva, “US 9,648,061 B2,” *United States Patent Office,* (2017), https://patentimages.storage.googleapis.com/40/d5/ed/2475e596d61e9a/US9648061.pdf
[64] Mark Andrejevic and Neil Selwyn, “Facial Recognition Technology in Schools: Critical Questions and Concerns,” *Learning, Media and Technology*, 45:2, 115-128, (2020) DOI: 10.1080/17439884.2020.1686014;

Spencer Ackerman, “TSA Screening Program Risks Racial Profiling Amid Shaky Science – Study Says.” *The Guardian*, February 8, 2017, https://www.theguardian.com/us-news/2017/feb/08/tsa-screening-racial-religious-profiling-aclu-study

[65] Lisa Feldman Barrett, Ralph Adochs, and Stacy Marsella, “Emotional Expressions Reconsidered: Challenges to Inferring Emotion from Human Facial Movements,” *Psychological Science in the Public Interest* 20, no. 1 (July 2019): 1–68; Zhimin Chen and David Whitney, “Tracking the Affective State of Unseen Persons,” *Proceedings of the National Academy of Sciences* 116, no. 15 (2019): 7559-7564, doi:10.1073/pnas.1812250116; Rens Hoegen, et al., "Signals of Emotion Regulation in a Social Dilemma: Detection from Face and Context," *2019 8th International Conference on Affective Computing and Intelligent Interaction* (ACII), IEEE, 2019: 1-7, doi:10.1109/acii.2019.8925478.

[66] Michel Foucault, *The History of Sexuality*, Translated by Robert J. Hurley. Harmondsworth, Middlesex: Penguin, 1990/1976; Michel Foucault, *Society Must be Defended: Lectures at the College de France, 1975–1976*. New York: Picador, 2003/1976.

[67] Yilun Wang and Michal Kosinski, "Deep neural networks are more accurate than humans at detecting sexual orientation from facial images," *Journal of personality and social psychology* 114, no. 2 (2018): 246.; Agüera y Arcas, Blaise, Alexander Todorov, and Margaret Mitchell, "Do algorithms reveal sexual orientation or just expose our stereotypes?," *Medium*, January 11, 2018, https://medium.com/@blaisea/do-algorithms-reveal-sexual-orientation-or-just-expose-our-stereotypes-d998fafdf477.; Xiaolin Wu and Xi Zhang, "Automated inference on criminality using face images." arXiv preprint arXiv:1611.04135 (2016): 4038-4052.; y Arcas, Blaise Aguera, Margaret Mitchell, and Alexander Todorov. "Physiognomy's new clothes," *Medium,* May 6, 2017, https://medium.com/@blaisea/physiognomys-new-clothes-f2d4b59fdd6a.; Coalition for Critical Technology, “Abolish the #TechToPrisonPipeline: Crime prediction technology reproduces injustices and causes real harm,” *Medium*, June 23, 2020, https://medium.com/@CoalitionForCriticalTechnology/abolish-the-techtoprisonpipeline-9b5b14366b16.

[68] Michel Foucault, 1977. *Discipline and Punish: The Birth of the Prison*. New York: Random House

[69] Luke Stark, “Facial Recognition Is the Plutonium of AI.”

[70] Woodrow Hartzog and Evan Selinger, “Facial Recognition Is the Perfect Tool for Oppression,” *Medium*, August 2, 2018, https://medium.com/s/story/facial-recognition-is-the-perfect-tool-for-oppression-bc2a08f0fe66.

[71] David Leslie, “Understanding Artificial Intelligence Ethics and Safety: A Guide for the Responsible Design and Implementation of AI systems in the Public Sector,” *The Alan Turing Institute*, June 11, 2019, http://doi.org/10.5281/zenodo.3240529.

[72] Information Commissioner's Office and The Alan Turing Institute, “Explaining Decisions Made with AI,” 2020, https://ico.org.uk/for-organisations/guide-to-data-protection/key-data-protection-themes/explaining-decisions-made-with-ai/.

[73] It is important to note here that permissible consent must be limited to reasonable expectations of the primary intended uses of facial images. There has been a vast abuse of individual consent in the internet scraping of personal pictures posted on social media and elsewhere for the compilation of largescale datasets as well as in the downstream “secondary” uses of these pictures as training resources for facial recognition systems. Aggressive technologists and developers have treated regimes of open access and re-use (for instance as directed by Creative Commons licenses) as a kind of blank check that has enabled blatant and systemic violation of privacy and consent rights. For a deeper dive into this and its related ethical consequences, see: Vinay Uday Prabhu and Abeba Birhane, "Large image datasets: A pyrrhic win for computer vision?"; Also the wider work on largescale datasets of MegaPixels: Harvey, Adam, and Jules

LaPlace. "MegaPixels: Origins, ethics, and privacy implications of publicly available face recognition image datasets." (2019), https://megapixels.cc

[74] See Woodrow Hartzog, "Privacy and the Dark Side of Control," *iai news*, September 4, 2017, https://iai.tv/articles/privacy-the-dark-side-of-control-auid-882.

[75] In the UK and European context this is largely a matter of abiding by human rights conventions and data protection laws. For instance the General Data Protection Regulation requires Data Protection Impact Assessments to be carried out in cases of high impact processing. These, "describe the nature, scope, context and purposes of the processing; assess necessity, proportionality and compliance measures; identify and assess risks to individuals; and identify any additional measures to mitigate those risks." See: Information Commissioner's Office, "Data protection impact assessments," 2020, https://ico.org.uk/for-organisations/guide-to-data-protection/guide-to-the-general-data-protection-regulation-gdpr/accountability-and-governance/data-protection-impact-assessments/; On the connection to human rights conventions, see also: European Union Agency for Fundamental Rights, "Facial recognition technology: fundamental rights considerations in the context of law enforcement," 2020, https://fra.europa.eu/sites/default/files/fra_uploads/fra-2019-facial-recognition-technology-focus-paper-1_en.pdf.

[76] Kak, ed., "Regulating Biometrics: Global Approaches and Urgent Questions," 26.

[77] Erik Learned-Miller, Vicente Ordóñez, Jamie Morgenstern, and Joy Buolamwini, "Facial Recognition Technologies in the Wild: A Call for a Federal Office," *Algorithmic Justice League*, (2020): 7-8.

## Figures Work Cited

**Acknowledgements:** This rapidly generated report would simply not exist without the heroic efforts of Morgan Briggs, who put together all of the figures and tables, edited content, and crafted the layout and design of the final product. Cosmina Dorobantu, Josh Cowls and Christopher Burr all shared helpful and incisive comments on previous drafts. Needless to say, all of the thoughts that are contained herein as well as any remaining errors are the responsibility of the author alone. Last but not least, a debt of gratitude is due to Corianna Moffatt for her stalwart bibiliographic and copyediting support.

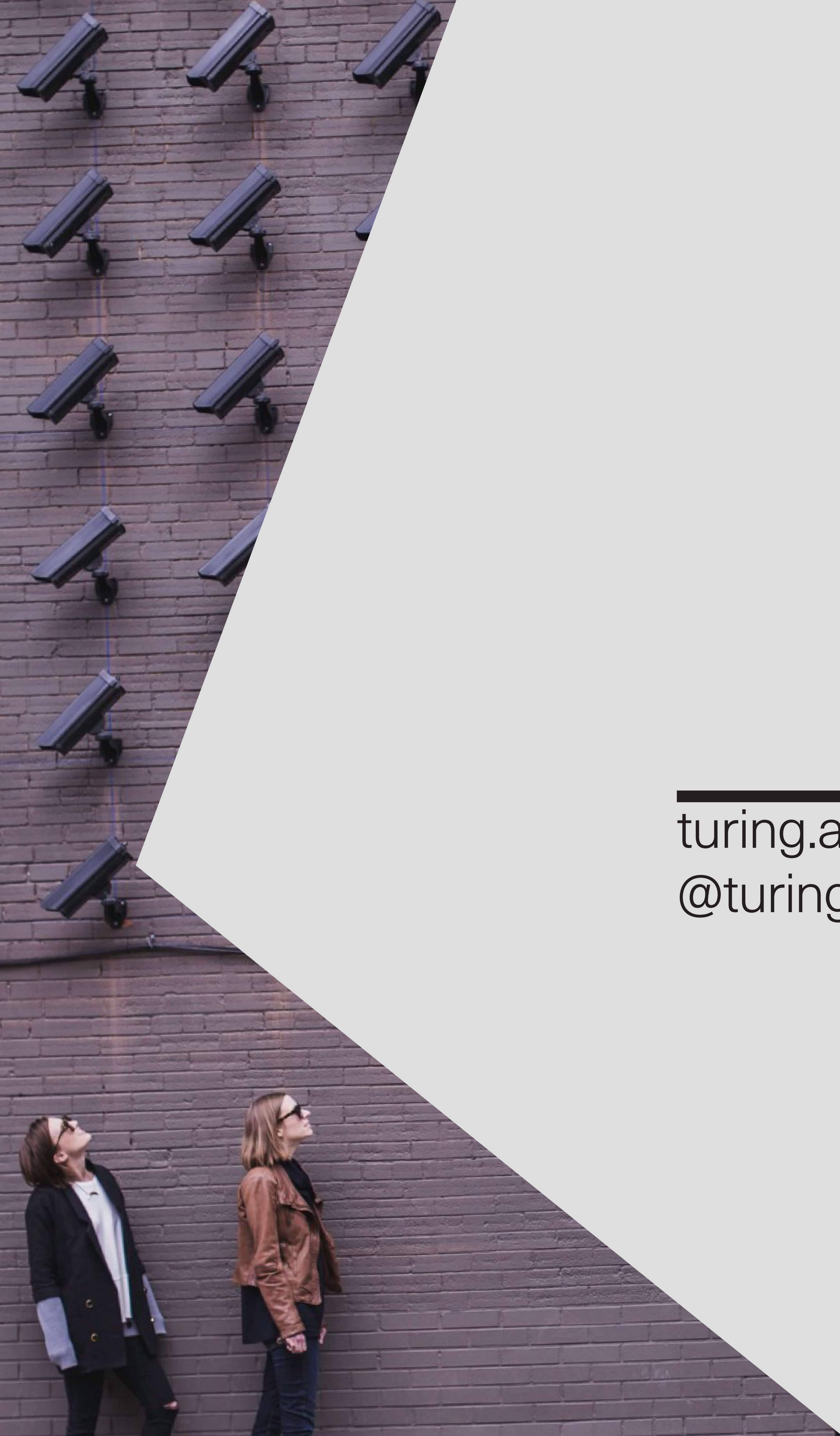

turing.ac.uk
@turinginst